\documentclass[10pt, conference]{IEEEtran}
\usepackage{cite}
\usepackage{bm}
\usepackage[cmex10]{amsmath}
\usepackage{nccmath}
\usepackage{amssymb,amsfonts}
\usepackage{algorithmic}
\usepackage{graphicx}
\usepackage{textcomp}
\usepackage{xcolor}
\usepackage{todonotes}
\usepackage{nicefrac}
\usepackage{xfrac}
\usepackage[font=small]{caption}
\usepackage[font=small]{subcaption}
\def\BibTeX{{\rm B\kern-.05em{\sc i\kern-.025em b}\kern-.08em
    T\kern-.1667em\lower.7ex\hbox{E}\kern-.125emX}}
\begin{document}

\title{\Large \textbf{What is the value of experimentation \& measurement?}
\\
{\normalsize Quantifying the value of reducing uncertainty to make better decisions}
}

\author{
\IEEEauthorblockN{C. H. Bryan Liu}
\IEEEauthorblockA{%
\textit{ASOS.com}, UK\\
bryan.liu@asos.com\vspace*{-20pt}}
\and
\IEEEauthorblockN{Benjamin Paul Chamberlain\footnote{work carried out while employed at ASOS.com}}
\IEEEauthorblockA{%
\textit{Twitter Inc}, UK\\
bchamberlain@twitter.com}\vspace*{-20pt}}

\maketitle

\begin{abstract}
Experimentation and Measurement (E\&M) capabilities allow organizations to accurately assess the impact of new propositions and to experiment with many variants of existing products. However, until now, the question of measuring the measurer, or valuing the contribution of an E\&M capability to organizational success has not been addressed. We tackle this problem by analyzing how, by decreasing estimation uncertainty, E\&M platforms allow for better prioritization. We quantify this benefit in terms of expected relative improvement in the performance of all new propositions and provide guidance for how much an E\&M capability is worth and when organizations should invest in one.
\end{abstract}


\section{Introduction}
\label{sec:introduction}
The value of making data driven or data informed decisions has become increasingly clear in recent years.
Key to making data driven decisions is the ability to accurately measure the impact of a given choice and to experiment with possible alternatives. 
We define Experimentation \& Measurement (E\&M) capabilities as the knowledge and tools necessary to run experiments (controlled or otherwise) with different products, services, or experiences, and measure their impact. The capabilities may be in the form of an online controlled experiment framework, a team of analysts, or a system capable of performing machine learning-aided causal inference.

The value of E\&M is currently best reflected in the success of major organizations that have adopted and advocated for them in the past decade. 
A large number of major technology companies report having mature infrastructure for online controlled experiments (OCEs, e.g. Google~\cite{tang10overlapping}, Linkedin~\cite{xu15frominfrastructure}, and Microsoft~\cite{kohavi13online}) and/or are heavily investing in state-of-the-art techniques (e.g. Airbnb~\cite{lee2018winner},
 Netflix~\cite{xie16improving}, and Yandex~\cite{poyarkov16boosted}).
Amazon~\cite{hill17efficient} and Facebook~\cite{gordon19comparison} have also reported the use of various causal inference techniques to measure the incrementality of advertising campaigns. A number of start-ups (e.g. Optimizely~\cite{johari17peeking} and Qubit~\cite{browne17whatworks}) have also recently been
established purely to manage OCEs for
businesses.

While mature E\&M capabilities can quantify the value of a proposition, it remains a major challenge to ``measure the measurer'' --- to quantify the value of the capabilities themselves. To the best of our knowledge, there is no work that addresses the question ``should we invest in E\&M capabilities" or how to value these capabilities, making it difficult to build a compelling business case to justify investment in the related personnel and infrastructure. We address this problem, calculating both the expected value and the risk, allowing the Sharpe ratio~\cite{sharpe1966mutual} for an E\&M capability to be calculated and compared to other potential investments.

The value created by E\&M capabilities can be divided into three classes --- 1) recognizing value 2) prioritizing propositions 3) optimizing individual propositions:

\textbf{1) Recognizing value} E\&M capabilities enable value to be attributed to a product, proposition or service. They also prevent damage from propositions that have negative value. This is important for dynamic organizations with large numbers of propositions as the damage caused  by individual roll outs can be compartmentalized and contained in a similar fashion to unit and integration testing in software development.

\textbf{2) Prioritization} Without E\&M capabilities, prioritization is based on back-of-envelope estimates or gut feel, which has high uncertainty. E\&M reduces the magnitude of the noise arising from estimation, enabling prioritization based on estimates that are closer to the true values and improved long-term decision making.

\textbf{3) Optimization} E\&M capabilities allow large numbers of variants to be evaluated against each other and the best to be selected efficiently. Without such capabilities, propositions can be experimented with sequentially, but this is slow and introduces noise from the changing environment.

While quantifying the values of 1) and 3) are relatively straightforward,\footnote{
The value of 1) comes from rolling back negative propositions. Given an E\&M capability, it can be calculated by summing the negative contributions of unsuccessful propositions. In the absence of a capability it can be estimated from the value distribution of propositions, which is given across industries in~\cite{johnson2017online} and~\cite{browne17whatworks}.
The value of 3) is the difference between the maximum and the mean value for each variant summed over the number of propositions. This can be estimated by placing Gaussian distributions over variants for each proposition or evaluated in the case that an E\&M capability exists.} quantifying the value of 2) is more interesting and the subject of the remainder of this paper. E\&M capabilities improve prioritization by reducing uncertainty in the value estimates of each proposition. This is a form of ranking under uncertainty, a well studied problem in the fields of statistics and operational research. However, in all previous work, either the variance is  assumed to be a fixed constant, or it is changed without the value being measured. Here we wish to understand the value of variance reduction through E\&M.


Our contribution is as follows. We 1) specify the first model that values the contribution of an E\&M capability in terms of better prioritization due to reduced estimation noise for propositions (Section~\ref{sec:mathematical_formulation}); 2) derive the variance of our estimate, allowing a Sharpe ratio to be calculated to guide organizations considering investment in E\&M (Section~\ref{sec:normal_normal_model}); and finally 3) provide two case studies based on large-scale meta-analyses that reflect how our model can be applied to real world practice (Section~\ref{sec:case_study}).


\section{Related work}
\label{sec:related_work}

There is a large literature on the use of controlled or natural experiments. A number of works are dedicated to running trustworthy online controlled experiments~\cite{dmitriev17adirtydozen}, choosing good metrics~\cite{hohnhold15focusing} and designing experiments where samples are dependent due to external confounders~\cite{backstrom11network,bakshy13uncertainty}. While important contributions, these works assume the existence of E\&M capabilities. However, to the best of our knowledge, there is no literature that helps organizations justify the acquisition of E\&M capabilities. We believe that filling this gap is necessary for wider adoption, and that increased participation will accelerate the development of the field.

This paper is related to existing work in statistics and operations research, in particular on decision making under uncertainty, which has been extensively studied since the 1980s. Notable work includes proposals for additional components in a decision maker's utility function~\cite{bell82regret}, alternate risk measures~\cite{xu09alternativerisk}, and a general framework for decision making with incomplete information (i.e. uncertainty)~\cite{weber87decision}. These works assume the inability to change the noise associated with estimation and/or measurement.

The sub-problem of ranking under uncertainty has also attracted considerable attention, partially due to the advent of large databases and the requirement in ranking results with certain ambiguity in relevance~\cite{soliman09ranking}. While Zuk et al.~\cite{zuk07ranking} measured the influence of noise levels in their work, they focused on the quality of the ranks themselves but not the value associated with the ranks.

The project selection problem is a related problem in optimization, where the goal is to find the optimal set of propositions using mixed integer linear programming, possibly under uncertainty. Work in this domain generally seeks methods that cope with existing risk/noise~\cite{mavrotas2013trichotomic}, and to the best of our knowledge there are no work that consider the value from reducing risk. While Shakhsi-Niaei et al.~\cite{shakhsiniaei11comprehensive} have discussed lowering the uncertainty level during the selection process, they refer to the uncertainty of decision parameters instead of the general noise level.

\begin{figure}
\vspace*{-4pt}
\begin{center}
\includegraphics[width=0.14\textwidth, trim = 0 0 0 0, clip]{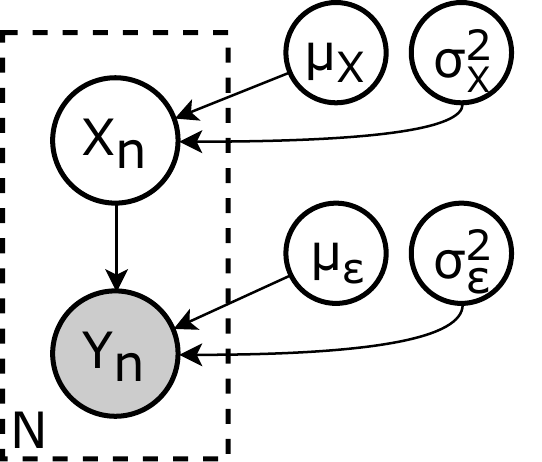}
\end{center}
\vspace*{-8pt}
\caption{The normal value / normal noise model for the value of propositions in plate notation. The $X_n$ represents the true, unobserved values and $Y_n$ represents the observed estimated values of propositions under some estimation noise.}
\label{fig:normal_normal_model_plate}
\vspace*{-12pt}
\end{figure}

\section{Mathematical Formulation}
\label{sec:mathematical_formulation}

We formulate the prioritization problem, and the value gained from E\&M capabilities, by considering $M$ propositions that must be selected from $N$ candidates, where $M<N$. 
The \textit{estimated} value of each proposition is given by $Y_n = X_n + \epsilon_n$, where~$X_n$ are the \textit{true} (unobserved) values that are estimated with error~$\epsilon_n$. 
The propositions are labelled in ascending order of estimated value $Y_n$ to get the order statistics ${Y_{(1)}, Y_{(2)}, ..., Y_{(N)}}$,
and the $M$ proposition with the highest estimated values: ${Y_{(N-M+1)}, Y_{(N-M+2)}, ..., Y_{(N)}}$ are selected. 
We are interested in the true value of the selected propositions, given by:
\begin{align}
    X_{\mathcal{I}(N-M+1)}, X_{\mathcal{I}(N-M+2)}, ..., X_{\mathcal{I}(N)},
    \label{eq:corresponding_true_val}
\end{align}
where $\mathcal{I}(\cdot)$ denotes the index function that maps the ranking to the index of the proposition.\footnote{Not to be confused with the set $X_{(N-M+1)}, X_{(N-M+2)}, ..., X_{(N)}$, which denotes the top $M$ propositions by their true value and are likely to be different than the set in~\eqref{eq:corresponding_true_val} \cite{zuk07ranking}.}

We define the mean true value of the $M$ selected propositions as
\begin{align}
    V \triangleq \nicefrac{1}{M}\left(X_{\mathcal{I}(N-M+1)} + X_{\mathcal{I}(N-M+2)} + ... + X_{\mathcal{I}(N)}\right),
    \label{eq:mean_true_value}
\end{align}
where a good prioritization maximizes $V$. Part of the value of E\&M capabilities arises from the observation that $V$ increases when the magnitude of the uncertainties arising from estimation ($\epsilon_n$) decreases.  We are interested in the value gained by reducing estimation uncertainty \textbf{without changing the set of propositions} (i.e. retaining all $X_n$s), as the true value of the propositions do not depend on the measurement method used:
\begin{align}
    D \triangleq V\, |_{\textrm{lower noise}} - V\, |_{\textrm{higher noise}} \,.
    \label{eq:D_pseudo}
\end{align}

We will derive the expected value of $D$ under various model assumptions in the following sections. Where applicable, we also provide bounds for the measure.

\section{Valuation Under Indep. Gaussian Assumptions}
\label{sec:normal_normal_model}

To value an E\&M capability, which is a generic framework that can be applied in many different ways across diverse organizations, it is first necessary to make some simplifying assumptions about the statistical properties of the propositions under consideration. To do this, we make use of~\cite{browne17whatworks} and~\cite{johnson2017online} whose authors performed a meta-analysis on the results of~6,700 e-commerce and~432 marketing experiments respectively. The uplifts indicated by the experiments, and hence the value of the propositions (under some estimation noise) exhibit the following properties:
\begin{enumerate}
    \item They can be positive or negative,
    \item They are usually clustered around an average instead of uniformly spreading across a certain range, and
    \item The distributions are heavy tailed. 
\end{enumerate}

We initially relax property and consider the case where the value of the propositions and the estimation noises are modelled with normal distributions:
\begin{align}
    X_n \overset{\textrm{i.i.d.}}{\sim} \mathcal{N}(\mu_X, \sigma^2_X)\;,\;
    \epsilon_n \overset{\textrm{i.i.d.}}{\sim} \mathcal{N}(\mu_\epsilon, \sigma^2_\epsilon),
\end{align}
where 
$\epsilon_n \perp X_m\, \forall\, n, m$ (see Figure~\ref{fig:normal_normal_model_plate}). This enables one to draw on the wealth of results in order statistics and Bayesian inference related to Gaussian distributions to begin with.\footnote{We extend the model to cover property 3 in the supplementary material\footnotemark[4] by modelling the value and noises by $t$-distributions, which do not have conjugate priors and hence introduce additional complexities. Empirical results show that the value gained under such assumption has a higher mean and variance, showing that the model can capture the ``higher risk, higher reward'' concept.}

We will derive the expected value and variance for~$V$, the mean true value of the top~$M$ propositions selected after being ranked by their estimated value (as defined in~\eqref{eq:mean_true_value}), as well as the expected value and an upper bound for the variance of~$D$, the value gained when the estimation noise is reduced. 
For brevity, we only present the derived quantities with key derivation steps in this paper. The full derivation is available in a supplemantary document.\footnote{Available on: https://github.com/liuchbryan/value\_of\_experimentation}

We will also demonstrate two key results. Firstly,
the expected mean true value of the selected propositions~($V$) increases when the estimation noise decreases, and the relative increase in value is dependent on how much noise we can reduce. Secondly, when $M$ is small, reducing the estimation noise may not lead to a statistically significant improvement in the true value of the propositions selected. As a result, improvements in prioritization driven by E\&M may only be justified for larger organizations.

\subsection{Calculating the Expectation}
\label{sec:gauss_mean}

We first derive the expected value for $D$. This requires the expected values of, in order:
\begin{enumerate}
    \item $Y_{(r)}$ - the \emph{estimated} value of the $r^{\textrm{th}}$ proposition, ranked in increasing estimated value;\footnote{Note $Y_{(r)}$ is equivalent to $Y_{\mathcal{I}(r)}$, as the propositions are ranked by their estimated values.}
    \item $X_{\mathcal{I}(r)}$ - the \emph{true} value of the $r^{\textrm{th}}$ proposition, ranked by increasing estimated value; and
    \item $V$ - the mean of the \emph{true} value for the $M$ most valuable propositions, ranked by their estimated values.
\end{enumerate}

To obtain the expected value for $Y_{(r)}$, we begin by observing that the $Y_n \overset{\textrm{i.i.d.}}{\sim} \mathcal{N}(\mu_X + \mu_\epsilon,\, \sigma^2_X + \sigma^2_\epsilon)$. We then apply a result by Blom \cite{blom1958statistical}, which states that the expected value for normal order statistics $Y_{(r)}$ can be closely approximated as:
\begin{align}
    \mathbb{E}(Y_{(r)}) \approx 
    \mu_X + \mu_\epsilon + 
    \sqrt{\sigma^2_X + \sigma^2_\epsilon}\;\Phi^{-1}\Big(\frac{r - \alpha}{N - 2\alpha + 1}\Big),
    \label{eq:observed_val_normal_order_stat_approx}
\end{align}
where $\Phi^{-1}$ denotes the quantile function of a standard normal distribution, and $\alpha\approx 0.4$ is a constant.\footnote{Decreasing the estimation noise $\sigma^2_\epsilon$ will decrease $\mathbb{E}(Y_{(r)})$ for any $r > \frac{N+1}{2}$, appearing to lower the average value of the top~$M$ propositions. This is a common pitfall; the estimated value of a proposition is not being optimized, what actually matters is the true, yet unobserved value of that proposition.}

The expected value of $X_{\mathcal{I}(r)}$ is obtained as follows. We first recall a standard result in Bayesian inference, which states that the posterior distribution of $X_n$ once $Y_n$ is observed is also normally distributed, with mean and variance given~by:
\begin{align}
    \mu_{X_n | \left (Y_n=y \right)} & = \frac{\sigma^2_X}{\sigma^2_X + \sigma^2_\epsilon} (y - \mu_\epsilon) + \frac{\sigma^2_\epsilon}{\sigma^2_X + \sigma^2_\epsilon} \mu_X 
    \label{eq:true_val_posterior_mean},
    \\
    \sigma^2_{X_n | \left (Y_n=y \right)} & = \sigma^2_\epsilon \sigma^2_X / (\sigma^2_X + \sigma^2_\epsilon) \,.
    \label{eq:true_val_posterior_variance}
\end{align}
We then apply the law of iterated expectations to obtain\footnotemark[5]
\begin{align}
    & \mathbb{E}(X_{\mathcal{I}(r)}) = \mathbb{E}\left(\mathbb{E}(X_{\mathcal{I}(r)} \,|\, Y_{(r)})\right)  \nonumber \\ 
    \approx & \, \mu_X + 
    \frac{\sigma^2_X}{\sqrt{\sigma^2_X + \sigma^2_\epsilon}}
   \;\Phi^{-1}\Big(\frac{r - \alpha}{N - 2\alpha + 1}\Big).
    \label{eq:true_val}
\end{align}

Equation~\eqref{eq:true_val} shows that decreasing the estimation noise $\sigma^2_\epsilon$ will lead to an increase in $\mathbb{E}(X_{\mathcal{I}(r)})$ for any $r > \frac{N+1}{2}$. It follows that the mean true value of the top $M$ propositions, selected according to their estimated value, will increase with the presence of a lower estimation noise. We show this by applying the expectation function to $V$ defined in~\eqref{eq:mean_true_value} to obtain
\begin{align}
    \mathbb{E}(V) \!\approx\mkern-1.5mu
    \mu_X \!+\! \frac{\sigma^2_X}{\sqrt{\mkern-.5mu\sigma^2_X \!+\! \sigma^2_\epsilon}} \mkern.5mu \frac{1}{M} \!\textstyle\sum\limits_{r=N\mkern-.5mu-\mkern-.5mu M\mkern-.5mu+\mkern-.5mu1}^{N} \!\! \Phi^{-1}\Big(\displaystyle\frac{r-\alpha}{N \!-\! 2\alpha \!+\! 1}\Big). \!
    \label{eq:expected_mean_true_value_simplified}
\end{align}
Note the complete absence of $\mu_\epsilon$ in this question, which suggests that systematic bias in estimation will not affect the true value of the chosen propositions under this process.

We finally consider the improvement when we reduce the estimation noise from $\sigma^2_\epsilon = \sigma^2_1$ to $\sigma^2_2$. This will be the expected value gained by having better E\&M capabilities:
\begin{align}
    & \mathbb{E}(D) = \mathbb{E}(V|_{\sigma^2_\epsilon = \sigma^2_2}) - \mathbb{E}(V|_{\sigma^2_\epsilon = \sigma^2_1}) \label{eq:expected_mean_true_value_diff}\\
    \approx & \Big(\frac{\sigma^2_X}{\sqrt{\sigma^2_X \!+\! \sigma^2_2}} \!-\! \frac{\sigma^2_X}{\sqrt{\sigma^2_X \!+\! \sigma^2_1}} \Big) \frac{1}{M} \textstyle\sum\limits_{r=N-M+1}^{N} \!\!\Phi^{-1}\Big(\displaystyle\frac{r-\alpha}{N \!-\! 2\alpha \!+\! 1}\Big).
    \nonumber
\end{align}
If we assume $\mu_X = 0$ (i.e. the true value of the propositions are centred around zero), then the relative gain is entirely dependent on $\sigma^2_X$, $\sigma^2_1$, $\sigma^2_2$:
\begin{align}
    & \frac{\mathbb{E}(D|_{\mu_X = 0})}{\mathbb{E}(V|_{\sigma^2_\epsilon = \sigma^2_1,\, \mu_X = 0})} 
    = \, \frac{\sqrt{\sigma^2_X + \sigma^2_1}}{\sqrt{\sigma^2_X + \sigma^2_2}} - 1 \,.
\label{eq:gauss_res}
\end{align}

To calculate the relative improvement in prioritization delivered by E\&M under these assumptions, plug into Equation~\eqref{eq:gauss_res}: 1) the estimated spread of the values ($\sigma^2_X$), 2) the estimated deviation of the current estimation process ($\sigma^2_1$), and 3) the estimated deviation to the actual value upon acquisition of E\&M capabilities ($\sigma^2_2$) to get an estimate on how much they will gain from acquiring such capabilities. 

\subsection{Calculating the Variance}
\label{sec:var_D}
To make effective investment decisions it is important to understand both the expected value and the risk or uncertainty that this value is delivered.
Having derived the expected value in~\eqref{eq:expected_mean_true_value_diff} and \eqref{eq:gauss_res}, in this section we address the investment risk given by the variance of $D$. Deriving the variance is similar to deriving the expectation --- one has to obtain the variances for (in order) $Y_{(r)}$, $X_{\mathcal{I}(r)}$, and $V$.
For the variance of $Y_{(r)}$, we apply a result from David and Johnson~\cite{david54statitical}, which states $\textrm{Var} \left ( Y_{(r)} \right )$ can be approximated as:
\begin{align}
    \textrm{Var}(Y_{(r)}) \approx
    \frac{r(N-r+1)}{(N+1)^2 (N+2)}
    \frac{\sigma^2_X + \sigma^2_\epsilon}{\big(\phi\big(\Phi^{-1}\big(\frac{r}{N+1}\big)\big)\big)^2} \,,
    \label{eq:observed_val_normal_order_stat_variance_approx}
\end{align}
where $\phi$ is the probability density function, and $\Phi^{-1}$ is the quantile function of a standard normal distribution.

The variance for $X_{\mathcal{I}(r)}$ is then obtained using the law of total variance:\footnotemark[5]
\begin{align}
    & \textrm{Var}(X_{\mathcal{I}(r)}) 
    =
    \mathbb{E}\big(\textrm{Var}(X_{\mathcal{I}(r)}|Y_{(r)})\big) +
    \textrm{Var}\big(\mathbb{E}(X_{\mathcal{I}(r)}|Y_{(r)})\big) \nonumber\\
    \approx\, &  
    \frac{\sigma^2_\epsilon \sigma^2_X}{\sigma^2_X + \sigma^2_\epsilon} + 
    \frac{\sigma^4_X}{\sigma^2_X + \sigma^2_\epsilon} \frac{r(N-r+1)}{(N+1)^2 (N+2)}
    \frac{1}{\big(\phi\big(\Phi^{-1}(\frac{r}{N+1})\big)\big)^2} \,.
    \label{eq:var_xir}
\end{align}

Before we derive the variance of $V$, we require the covariance between pairs of~$Y_{(\cdot)}$s and~$X_{\mathcal{I}(\cdot)}$s. This is necessary as the terms of $V$ (see~\eqref{eq:mean_true_value}), being the result of removing noise from successive order statistics, are highly correlated.

David and Nagaraja \cite{david2004order} have provided a formula to estimate the covariance between~$Y_{(r)}$ and~$Y_{(s)}$ for any $r,s\leq N$:
\begin{align}
    & \textrm{Cov}(Y_{(r)}, Y_{(s)}) \nonumber\\ \approx \,&
    \frac{r(N-s+1)}{(N+1)^2(N+2)}
    \frac{\sigma^2_X + \sigma^2_\epsilon}{\phi\big(\Phi^{-1}(\frac{r}{N+1})\big)\,\phi\big(\Phi^{-1}(\frac{s}{N+1})\big)} \,.
    \label{eq:cov_yr_ys}
\end{align}

To obtain the covariance between $X_{\mathcal{I}(r)}$ and $X_{\mathcal{I}(s)}$ for any $r, s \leq N$, we use the law of total covariance with multiple conditioning variables \cite{bowsher12identifying} to obtain\footnotemark[5]
\begin{align}
    & \textrm{Cov}(X_{\mathcal{I}(r)}, X_{\mathcal{I}(s)}) \label{eq:cov_xir_xis_final}\\
    = \, & \mathbb{E}\big(\mathbb{E}(\textrm{Cov}(X_{\mathcal{I}(r)}, X_{\mathcal{I}(s)} | Y_{(r)}, Y_{(s)}) | Y_{(r)})\big) + \nonumber \\
    & \!\!\! \mathbb{E}\big(\textrm{Cov}(\mathbb{E}(X_{\mathcal{I}(r)} | Y_{(r)}, Y_{(s)}), \mathbb{E}(X_{\mathcal{I}(s)} | Y_{(r)}, Y_{(s)}) | Y_{(r)} )\big) \, + \nonumber \\
    & \!\!\! \textrm{Cov}\big(\mathbb{E}(\mathbb{E}(X_{\mathcal{I}(r)} | Y_{(r)}, Y_{(s)}) | Y_{(r)}), \mathbb{E}(\mathbb{E}(X_{\mathcal{I}(s)} | Y_{(r)}, Y_{(s)}) | Y_{(r)})\big)
    \nonumber\\
    \approx & 
    \frac{\sigma^4_X}{\sigma^2_X + \sigma^2_\epsilon} \frac{r(N-s+1)}{(N+1)^2(N+2)}\frac{1}{\phi\big(\Phi^{-1}(\frac{r}{N+1})\big)\,\phi\big(\Phi^{-1}(\frac{s}{N+1})\big)}. \nonumber
\end{align}
Equation~\eqref{eq:cov_xir_xis_final} affirms the claim that the $X_{\mathcal{I}(\cdot)}$s are positively correlated. Unlike the $X_n$s, which are independent by definition, they become correlated under the presence of ranking information.
Now we can state the variance of $V$ and $D$. Applying the variance function to~\eqref{eq:mean_true_value} we get
\begin{align}
    & \textrm{Var}(V)
    = \frac{1}{M^2}\big(\textstyle\sum\nolimits_{r=N-M+1}^{N} \textrm{Var}\left(X_{\mathcal{I}(r)}\right) \, + \nonumber\\
    & \quad\;\; \textstyle\sum\nolimits_{r=N-M+1}^{N} \, \textstyle\sum\nolimits_{s=r+1}^{N} 2 \cdot \textrm{Cov}\left(X_{\mathcal{I}(r)}, X_{\mathcal{I}(s)}\right)\big) ,
    \label{eq:var_V}
\end{align}
where $\textrm{Var}(X_{\mathcal{I}(r)})$ and $\textrm{Cov}(X_{\mathcal{I}(r)}, X_{\mathcal{I}(s)})$ are defined in~\eqref{eq:var_xir} and \eqref{eq:cov_xir_xis_final}.
The variance of $D$ is thus:
\begin{align}
    & \textrm{Var}(D) = \textrm{Var}(V |_{\sigma^2_\epsilon = \sigma^2_2} - V |_{\sigma^2_\epsilon = \sigma^2_1}) \label{eq:var_D}\\
    = & \textrm{Var}\big(V |_{\sigma^2_\epsilon = \sigma^2_2}\big) \!+ \textrm{Var}\big(V |_{\sigma^2_\epsilon = \sigma^2_1}\big) \!- 2 \, \textrm{Cov}\big(V |_{\sigma^2_\epsilon = \sigma^2_2}, V |_{\sigma^2_\epsilon = \sigma^2_1}\big) . \nonumber
\end{align}
The first two terms on the RHS of~\eqref{eq:var_D} are that defined in~\eqref{eq:var_V}, while the last term can be expanded as follow:
\begin{align}
   & \textrm{Cov}\big(V |_{\sigma^2_\epsilon = \sigma^2_2}, V |_{\sigma^2_\epsilon = \sigma^2_1}\big) \label{eq:cov_v_diff_noise}\\
   = \, & \frac{1}{M^2} \textstyle\sum\limits_{r=N-M+1}^{N} \textstyle\sum\limits_{s=r}^{N} \, 2 \cdot \textrm{Cov}\big(X_{\mathcal{I}(r)} |_{\sigma^2_\epsilon = \sigma^2_2},\, X_{\mathcal{I}(s)} |_{\sigma^2_\epsilon = \sigma^2_1}\big) \,. \nonumber
\end{align}
Equation~\eqref{eq:cov_v_diff_noise} shows the covariance term in~\eqref{eq:var_D} is positive as all its components are positive (cf.~\eqref{eq:cov_xir_xis_final}, albeit with a different magnitude). Hence the variance terms in~\eqref{eq:var_D} form an upper bound to the variance of~$D$:
\begin{align}
     \textrm{Var}(D) < 
     \textrm{Var}\big(V |_{\sigma^2_\epsilon = \sigma^2_2}\big) + \textrm{Var}\big(V |_{\sigma^2_\epsilon = \sigma^2_1}\big) .
\end{align}
In practice, the variance of $D$ is much lower than the bound, due to the $V$s being highly correlated.

We conclude this section by observing that $M$ and $N$ have a large influence on $\textrm{Var}(D)$, appearing as squared terms (as opposed to $\sigma^2_1$ and $\sigma^2_2$, which are linear). This is crucial as even in cases where the $\mathbb{E}(D)$ is positive, the limited capacity of an organization to introduce new propositions may mean that the Sharpe ratio~\cite{sharpe1966mutual}, defined as 
\begin{align}
    (\mathbb{E}(D)-r) / \sqrt{\textrm{Var}(D)},
\end{align}
where $r$ is a small constant, may not be high enough to justify investment in an E\&M capability.

The exact threshold where an organization should consider acquiring such capabilities depends on multiple factors including their size (which affects $M$), the size of their backlog ($N$), the nature of their work ($\mu_X$ and $\sigma^2_X$), and how good they were at estimation ($\sigma^2_1$). We refrain from providing a one-size-fits-all recommendation, but give examples in Section~\ref{sec:case_study}.

\begin{figure*}
\vspace*{-7pt}
\begin{center}
\begin{subfigure}[t]{.153\textwidth}
  \includegraphics[width=0.97\textwidth, trim=3mm 0 2.8mm 0, clip]{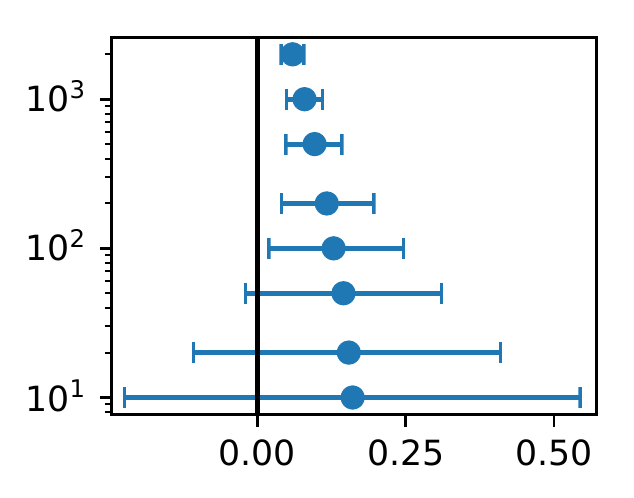}
  \vspace*{-6pt}
  \caption{$(1\%)^2, (0.8\%)^2$}
  \label{fig:value_gained_browne_johnson_1_0-8}
\end{subfigure}
\hspace*{.001\textwidth}
\begin{subfigure}[t]{.153\textwidth}
  \includegraphics[width=0.97\textwidth, trim=3mm 0 2.8mm 0, clip]{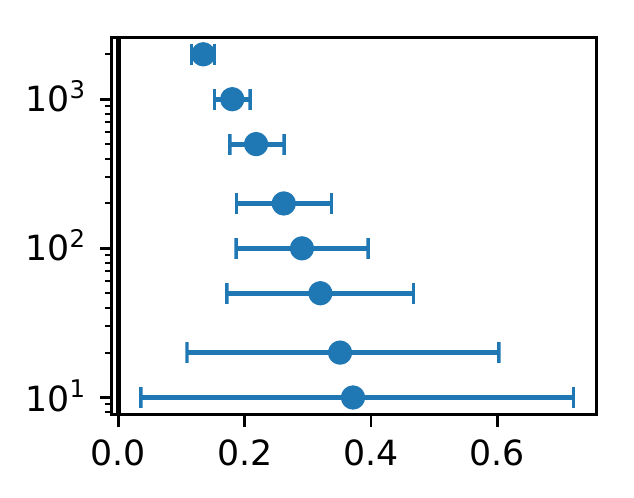}
  \vspace*{-6pt}
  \caption{$(1\%)^2, (0.6\%)^2$}
  \label{fig:value_gained_browne_johnson_1_0-6}
\end{subfigure}
\hspace*{.001\textwidth}
\begin{subfigure}[t]{.153\textwidth}
  \includegraphics[width=0.97\textwidth, trim=3mm 0 2.8mm 0, clip]{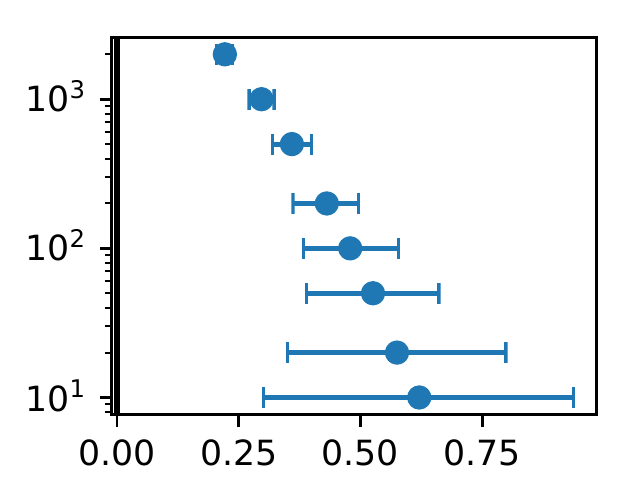}
  \vspace*{-6pt}
  \caption{$(1\%)^2, (0.4\%)^2$}
  \label{fig:value_gained_browne_johnson_1_0-4}
\end{subfigure}
\hspace*{.001\textwidth}
\begin{subfigure}[t]{.153\textwidth}
  \includegraphics[width=0.97\textwidth, trim=3mm 0 2.8mm 0, clip]{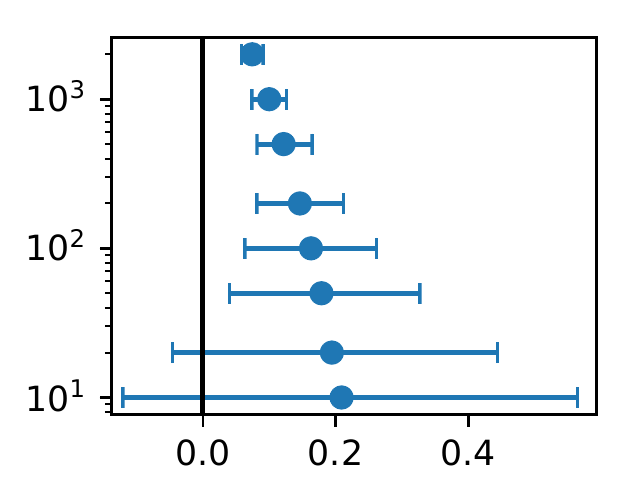}
  \vspace*{-6pt}
  \caption{$(0.8\%)^2, (0.6\%)^2$}
  \label{fig:value_gained_browne_johnson_0-8_0-6}
\end{subfigure}
\hspace*{.001\textwidth}
\begin{subfigure}[t]{.153\textwidth}
  \includegraphics[width=0.97\textwidth, trim=3mm 0 2.8mm 0, clip]{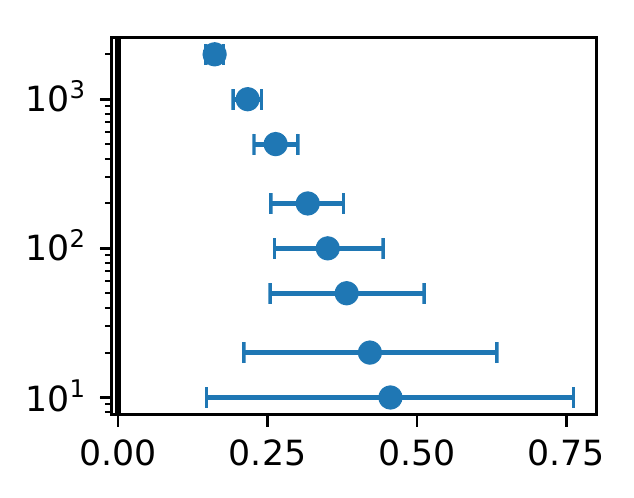}
  \vspace*{-6pt}
  \caption{$(0.8\%)^2, (0.4\%)^2$}
  \label{fig:value_gained_browne_johnson_0-8_0-4}
\end{subfigure}
\hspace*{.001\textwidth}
\begin{subfigure}[t]{.153\textwidth}
  \includegraphics[width=0.97\textwidth, trim=3mm 0 2.8mm 0, clip]{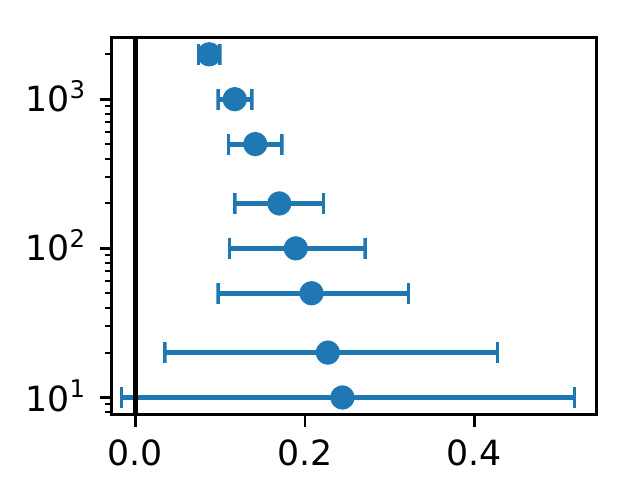}
  \vspace*{-6pt}
  \caption{$(0.6\%)^2, (0.4\%)^2$}
  \label{fig:value_gained_browne_johnson_0-6_0-4}
\end{subfigure}
\end{center}
\vspace*{-7pt}
\caption{The value gained by having some experimentation \& measurement (E\&M) capabilities (x-axis, in percent) under different capacity~$M$ ($y$-axis, in log scale) in the case study on 6700 e-commerce experiments reported by Browne and Johnson~\cite{browne17whatworks} (see Section~\ref{sec:case_study_ecommerce}). In each plot the dot represents the mean, and the error bar represents the 5th--95th percentile of the empirical value distribution. The subcaption denotes the estimation noise before \& after acquisition of E\&M capabilities (i.e. $\sigma^2_1, \sigma^2_2$). We fix $\mu_X, \mu_\epsilon = 0$, $\sigma^2_X = (0.7\%)^2$, and $N = 6700$.}
\label{fig:value_gained_browne_johnson}
\vspace*{-10pt}
\end{figure*}

\section{Experiments}
\label{sec:experiments}

Having performed theoretical calculations for the expectation and variance of the value E\&M systems deliver through enhanced prioritization, here we verify those calculations using simulation results. All code used in the experiments, case studies and extensions is available on GitHub.\footnotemark[4]


We first verify the result derived in Section~\ref{sec:normal_normal_model} empirically.
For each run, we fix the value of $N$, $M$, $\mu_X$, $\mu_\epsilon$, $\sigma^2_X$, $\sigma^2_1$ (the higher~$\sigma^2_\epsilon$), and $\sigma^2_2$ (the lower~$\sigma^2_\epsilon$). This is followed by 5,000 cycles of the same operations to obtain samples for $V$ and $D$:\footnote{Identifiers in \texttt{monospace} refer to variables used in software packages, which correspond to the random variables used in Section \ref{sec:normal_normal_model}.}
\begin{enumerate}
    \itemsep0.5mm
    \item Take $N$ samples from $\mathcal{N}(\mu_X, \sigma^2_X)$, referred as \texttt{X\textsubscript{n}} hereafter with \texttt{n} being the index; 
    \item Take $N$ samples from $\mathcal{N}(\mu_\epsilon, \sigma^2_1)$, and sum the \texttt{n}\textsuperscript{th}-indexed sample with \texttt{X\textsubscript{n}} $\forall$\texttt{n} to obtain \texttt{Y\textsubscript{n}}$|_{\sigma^2_\epsilon = \sigma^2_1}$;
    \item Rank the \texttt{Y\textsubscript{n}}s and obtain the indices of the $M$ largest samples;\footnote{These indices corresponds to the set $\{{\mathcal{I}(r)}\}_{r=N-M+1}^{N}$ in Section \ref{sec:normal_normal_model}.}
    \item Take the \texttt{X\textsubscript{n}}s where \texttt{n} is in the set of indices obtained in Step 3, and calculate the mean \texttt{V}$|_{\sigma^2_\epsilon = \sigma^2_1}$;
    \item Without replacing the \texttt{X\textsubscript{n}}s obtained in Step 1), repeat Steps 2 to 4 with $\sigma^2_\epsilon = \sigma^2_2$ (i.e. generate samples from $\mathcal{N}(\mu_\epsilon, \sigma^2_2)$ in Step 2) to obtain \texttt{V}$|_{\sigma^2_\epsilon = \sigma^2_2}$; and
    \item Take the difference between \texttt{V}$|_{\sigma^2_\epsilon = \sigma^2_2}$ obtained in Step~5 and \texttt{V}$|_{\sigma^2_\epsilon = \sigma^2_2}$ from Step 4 to get \texttt{D}.
\end{enumerate}

We expect the mean and variance of the samples obtained in Steps~4 and~5 to match the RHS of~\eqref{eq:expected_mean_true_value_simplified} and~\eqref{eq:var_V} respectively, and the mean of the samples obtained in Step~6 to match the RHS of~\eqref{eq:expected_mean_true_value_diff}. To verify this, we perform~1,000 bootstrap resamplings on the samples obtained above to obtain an empirical bootstrap distribution of the sample mean and variance in each run. The $(1-\alpha)\%$ bootstrap resampling confidence interval (BRCI) should then cntain the theoretical mean/variance $(1-\alpha)\%$ of the times.

We performed a total of 351 runs, using a set of parameters that are randomly chosen from a curated parameter space. We observed that the quantities $\mathbb{E}(V|_{\sigma^2_1})$, $\textrm{Var}(V|_{\sigma^2_1})$, $\mathbb{E}(V|_{\sigma^2_2})$, $\textrm{Var}(V|_{\sigma^2_2})$, and $\mathbb{E}(D)$ fall within the 95\% centered BRCI 336, 320, 336, 305, and 339 times respectively. While these numbers are expected for the expectations, they are on the low side for the variances. Upon further investigation we realized that the majority of the out-of-BRCI cases have a theoretical variance below the BRCI (25 below, six over for $\textrm{Var}(V|_{\sigma^2_1})$; and 39 below, seven over for $\textrm{Var}(V|_{\sigma^2_2})$), suggesting a slight underestimate in our variance derivation. We believe that this is due to the omission of higher order terms when using the formulas in~\cite{david54statitical}, leading to a fraction of a percent bias. The bias is more apparent when~$N$ and $M$ are small. Otherwise, we are satisfied with the soundness of the derived quantities.

\section{Case study}
\label{sec:case_study}

\emph{``What do e-commerce / marketing companies gain by acquiring experimentation \& measurement capabilities?''}

It is difficult to verify any model that seeks to ascertain the value of E\&M capabilities with real data. This is not only because of the inability to observe the true value of a proposition/product/service, but also the lack of published measurements from organizations. The closest proxies are meta-analyses, including that compiled by Browne and Johnson~\cite{browne17whatworks} and Johnson et al.~\cite{johnson2017online}, which contain statistics on the measured uplift (in relative \%) over a large number of e-commerce and marketing experiments for many organizations.

The information presented by the two groups of researchers are sufficient for us to ask the following question: If all the experiments presented by Browne and Johnson / Johnson et al. are conducted for the same organization, how much value did the E\&M capabilities add due to improved prioritization?

\begin{figure*}
\vspace*{-7pt}
\begin{center}
\begin{subfigure}[t]{.155\textwidth}
  \includegraphics[width=.97\textwidth, trim=3mm 0 2.8mm 0, clip]{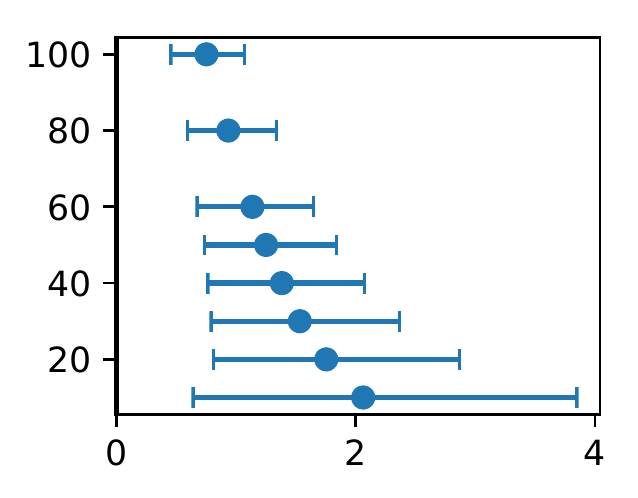}
  \vspace*{-6pt}
  \caption{$(5\%)^2, (0.8\%)^2$}
  \label{fig:value_gained_johnson_etal_5_0-8}
\end{subfigure}
\hspace*{.001\textwidth}
\begin{subfigure}[t]{.155\textwidth}
  \includegraphics[width=0.97\textwidth, trim=3mm 0 2.8mm 0, clip]{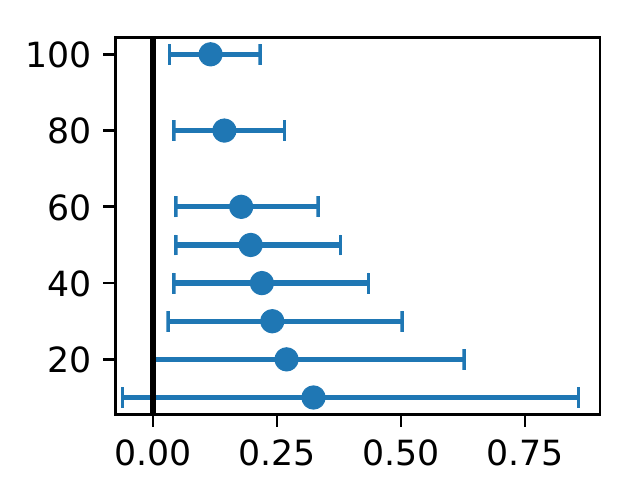}
  \vspace*{-6pt}
  \caption{$(2\%)^2, (0.8\%)^2$}
  \label{fig:value_gained_johnson_etal_2_0-8}
\end{subfigure}
\hspace*{.001\textwidth}
\begin{subfigure}[t]{.155\textwidth}
  \includegraphics[width=.97\textwidth, trim=3mm 0 2.8mm 0, clip]{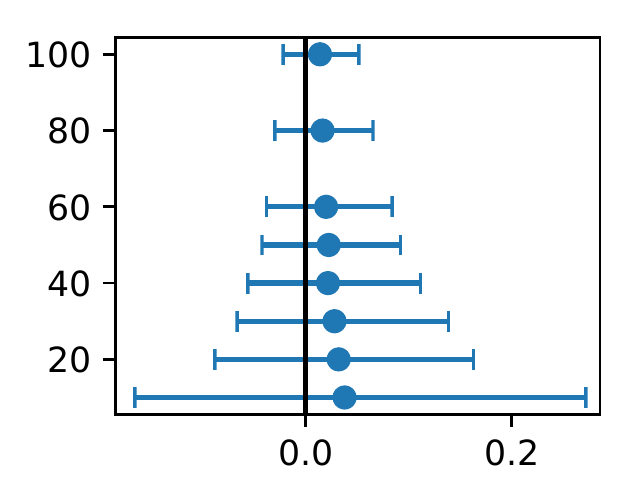}
  \vspace*{-6pt}
  \caption{$(1\%)^2, (0.8\%)^2$}
  \label{fig:value_gained_johnson_etal_1_0-8}
\end{subfigure}
\hspace*{.001\textwidth}
\begin{subfigure}[t]{.155\textwidth}
  \includegraphics[width=0.97\textwidth, trim=3mm 0 2.8mm 0, clip]{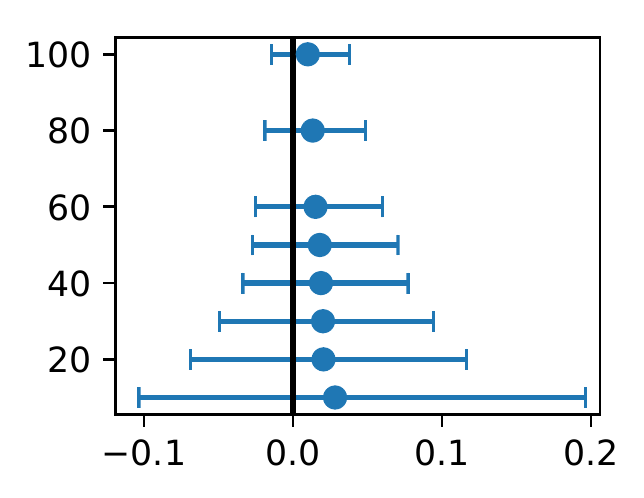}
  \vspace*{-6pt}
  \caption{$(0.8\%)^2, (0.6\%)^2$}
  \label{fig:value_gained_johnson_etal_0-8_0-6}
\end{subfigure}
\hspace*{.001\textwidth}
\begin{subfigure}[t]{.155\textwidth}
  \includegraphics[width=0.97\textwidth, trim=3mm 0 2.8mm 0, clip]{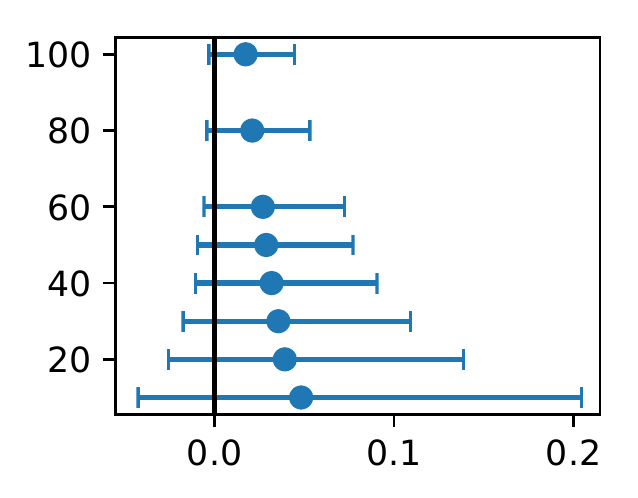}
  \vspace*{-6pt}
  \caption{$(0.8\%)^2, (0.4\%)^2$}
  \label{fig:value_gained_johnson_etal_0-8_0-4}
\end{subfigure}
\hspace*{.001\textwidth}
\begin{subfigure}[t]{.155\textwidth}
  \includegraphics[width=0.97\textwidth, trim=3mm 0 2.8mm 0, clip]{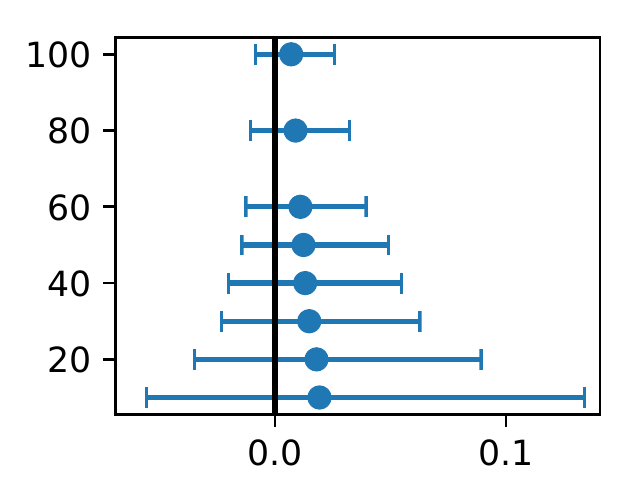}
  \vspace*{-6pt}
  \caption{$(0.6\%)^2, (0.4\%)^2$}
  \label{fig:value_gained_johnson_etal_0-6_0-4}
\end{subfigure}
\end{center}
\vspace*{-7pt}
\caption{The value gained by having some experimentation \& measurement (E\&M) capabilities (x-axis, in percent) under different capacity $M$ ($y$-axis) in the case study on 184 marketing experiments reported by Johnson et al.~\cite{johnson2017online} (see Section~\ref{sec:case_study_marketing}). In each plot the dot represents the mean, and the error bar represents the 5th-95th percentile of the empirical value distribution. The subcaption denotes the estimation noise before and after acquisition of E\&M capabilities (i.e. $\sigma^2_1, \sigma^2_2$). Here we fix $\mu_X = 19.9\%$, $\mu_\epsilon = 0$, $\sigma^2_X = (10\%)^2$, and $N = 184$.}
\label{fig:value_gained_johnson_etal}
\vspace*{-10pt}
\end{figure*}

\subsection{e-Commerce companies}
\label{sec:case_study_ecommerce}

In~\cite{browne17whatworks} Browne and Johnson reported running 6,700 A/B test in e-commerce companies, with an overall effect in relative ConVersion Rate (CVR) uplift centered at around zero, and the 5\% and 95\% percentiles at around $\pm [1.2\%, 1.3\%]$. We then divide the range by ${z_{0.95} \approx 1.645}$, the 95\textsuperscript{th} percentile of a standard normal, to estimate the distribution reported has a standard deviation of around 0.75\%.
Based on this information, we take $\mu_X = 0$ and $\sigma^2_X = (0.6\%)^2$ (taking into account that the reported distribution incorporated some estimation noise, and hence the spread of the true values should be slightly lower).

Given an A/B test on CVR uplift run by the largest organizations (e.g. one with five million visitors and a 5\% CVR) carries an estimation noise of around~$(0.28\%)^2$,\footnote{The estimation noise ($\sigma^2_\epsilon$) from an A/B test measuring conversion rate (CVR) uplift is the variance of the distribution on the difference in CVR between two variants under a no-difference null hypothesis. This equals to $2\cdot\frac{p(1-p)}{n}$, i.e. twice the variance of the sample CVR $p$ with $n$ samples.} we explore the scenarios where we reduce the noise level from ${\sigma^2_1 = \{(1\%)^2, (0.8\%)^2, (0.6\%)^2\}}$ to ${\sigma^2_2 = \{(0.8\%)^2, (0.6\%)^2, (0.4\%)^2\}}$, representing different levels of estimation abilities before and after acquisition of E\&M capabilities for companies of various sizes. We also calculate the value gained under different~$M$s (from 10 to 2000) to simulate organizations with different, yet realistic capacities, while fixing $N = 6700$ (\# experiments). We set $\mu_\epsilon = 0$ as we do not assume any systematic bias during estimation in this case.

Results are reported in Figure~\ref{fig:value_gained_browne_johnson}, which shows the relationship between different $M$s and the value gained under different magnitudes of estimation noise reduction. One can observe that the expected gain in value actually decreases in~$M$. This is expected: as one increases their capacity, they will run out of the most valuable work, and have to settle for less valuable work that has many acceptable replacements with similar value, limiting the value E\&M capabilities bring.

We can also see an inverse relation between the size of $M$ and the uncertainty of the value gained. As a result, while the expected value gain decreases with increasing $M$, the uncertainty drops quicker such that at some $M$ we will see a statistically significant increase in value gained, and/or an acceptable Sharpe ratio that justifies investment in E\&M capabilities. The specific value that tips the balance is heavily dependent on individual circumstances.

\subsection{Marketing companies}
\label{sec:case_study_marketing}

In the second case-study we repeat the process applied to e-commerce in Section~\ref{sec:case_study_ecommerce} for the marketing experiments described in~\cite{johnson2017online}. In that work Johnson et al. reported running~184 marketing experiments that measures CVR, with an mean relative uplift of 19.9\% and standard error of 10.8\%. This suggests the use of $\mu_X = 19.9\%$ and $\sigma^2_X = (10\%)^2$, the latter slightly reduced to account for the estimation noise being included in the reported standard error.

Johnson et al. also noted the average sample size in these experiments is over five million, which keeps the estimation noise low. However, the design of marketing experiments often comes with extra sources of noise compared to standard A/B tests~\cite{gordon19comparison,liu2018designing}, hence we keep the estimation noise in our scenarios the same as above (i.e. ${\sigma^2_2 = \{(0.8\%)^2, (0.6\%)^2, (0.4\%)^2\}}$). The larger variance in the uplifts provide room for us to assume a larger estimation error without E\&M capabilities, and we explore the scenario where ${\sigma^2_1 = \{(5\%)^2, (2\%)^2, (1\%)^2, (0.8\%)^2, (0.6\%)^2\}}$. We set~$N=184$ (\# experiments), and vary~$M$  between 10 and 100 for each combination of~$\sigma^2_1$ and~$\sigma^2_2$.

Figure~\ref{fig:value_gained_johnson_etal} shows the results. We can see in the presence of a larger variability in the true uplift of the advertising campaigns ($\sigma^2_X$) and lower capacity ($M$), the level of estimation noise reduction that gave a statistically significant value gained in the e-commerce example is no longer sufficient. One needs a larger noise reduction, or to increase their capacity to effectively control the risk in investing in E\&M capabilities. Otherwise they may be better off focusing their resources on improving their limited number of existing propositions.

\section{Conclusion}
\label{sec:conclusion}

We have addressed the problem of valuing E\&M capabilities. Such capabilities deliver three forms of value to organizations. These are 1) improved recognition of the value of propositions 2) enhanced capability to prioritize and 3) the ability to optimize individual propositions. Of these, the most challenging to address is improved prioritization. We have established a methodology to value better prioritization through reduced estimation error using the framework of ranking under uncertainty. The key insight is that E\&M capabilities reduce the estimation error in the value of individual propositions, allowing prioritization to follow more closely the optimal order of projects were the true values of propositions be observable. We have provided simple formulae that give the value of E\&M capabilities and the Sharpe ratio governing investment decisions and provide guidelines for conditions when such investments are not appropriate. 

\bibliographystyle{ieeetr}
\bibliography{testing-bibliography-short} 

\newpage
\begin{appendices}
\Large \noindent\textbf{Supplementary Document}

\vspace*{10pt}
\normalsize
This document is intended to be used as a supplement to the paper ``What is the value of experimentation \& measurement?" by Liu and Chamberlain. We first show in Appendix A a full derivation of the quantities presented in Section~\ref{sec:normal_normal_model} of the paper. This is followed by three empirical extensions to the model described in the section that open the door for future work in Appendix B.

\section{Full Derivation of Valuation Under Independent Gaussian Assumptions}

In this appendix we provide the full derivation of the quantities presented in Section~\ref{sec:normal_normal_model} of the paper. We begin by showing the standard Bayesian inference result (in one-dimensional form) quoted in Equations~\eqref{eq:true_val_posterior_mean} and~\eqref{eq:true_val_posterior_variance} in Section~\ref{sec:1d}. This is followed by a statement of the result in its multi-dimensional form, and the derivation of its specialisation  
used in Equation~\eqref{eq:cov_xir_xis_final} in Section~\ref{sec:2d}. We finally provide the full derivation of that presented in Sections~\ref{sec:gauss_mean} and~\ref{sec:var_D} in Sections~\ref{sec:appendix_gauss_mean} and~\ref{sec:appendix_var_D} respectively. 

\subsection{Mean \& variance of a conditioned normal r.v.}
\label{sec:1d}

We first replicate the setup in Section~\ref{sec:normal_normal_model} of the paper.
Let $X_n \overset{\textrm{i.i.d.}}{\sim} \mathcal{N}(\mu_X, \sigma^2_X)$ and $\epsilon_n \overset{\textrm{i.i.d.}}{\sim} \mathcal{N}(\mu_\epsilon, \sigma^2_\epsilon)$, where all $X_n$s and $\epsilon_n$s are mutually independent. In addition let ${Y_n = X_n + \epsilon_n}$, this means 
\begin{align}
    Y_n \sim \mathcal{N}(\mu_X + \mu_\epsilon\,,\, \sigma^2_X + \sigma^2_\epsilon)
\end{align} and the conditional distribution of $Y_n$ given $X_n = x$ is
\begin{align}
    (Y_n \,|\, X_n = x) \;\sim\; \mathcal{N}(x + \mu_\epsilon,\, \sigma^2_\epsilon).
\end{align}

A standard Bayesian inference result states the conditional distribution of $X_n$ given $Y_n = y$ is a normal distribution with
\begin{align}
    \mu_{X_n | (Y_n = y)} & = \frac{\sigma^2_X}{\sigma^2_X + \sigma^2_\epsilon} (y - \mu_\epsilon) + \frac{\sigma^2_\epsilon}{\sigma^2_X + \sigma^2_\epsilon} \mu_X ,
    \label{eq:appendix_posterior_mu}\\
    \sigma^2_{X_n | (Y_n = y)} & = \frac{\sigma^2_\epsilon \sigma^2_X}{\sigma^2_X + \sigma^2_\epsilon} \,.
    \label{eq:appendix_posterior_sigma_sq}
\end{align}

We will show below that this is indeed the case. We begin by specifying the conditional distribution $X_n\,|\,Y_n$ as the posterior distribution given $X_n$ as the prior and $Y_n\,|\,X_n$ as the likelihood:
\begin{align}
    & f_{X_n|Y_n} (x\,|\,y) 
      = \frac{f_{Y_n|X_n}(y\,|\,x)\, f_{X_n}(x)}{f_{Y_n}(y)} \nonumber\\
     = & \frac{\frac{1}{\sqrt{2\pi}\sqrt{\sigma^2_\epsilon}}\exp{\!\left(\!-\frac{1}{2}\frac{\left(y - (x+\mu_\epsilon)\right)^2}{\sigma^2_\epsilon}\right)}\frac{1}{\sqrt{2\pi}\sqrt{\sigma^2_X}}\exp{\!\left(\!-\frac{1}{2}\frac{\left(x - \mu_X \right)^2}{\sigma^2_X}\right)}}{\frac{1}{\sqrt{2\pi}\sqrt{\sigma^2_X + \sigma^2_\epsilon}}\exp{\left(-\frac{1}{2}\frac{\left(y - (\mu_X+\mu_\epsilon)\right)^2}{\sigma^2_X + \sigma^2_\epsilon}\right)}} .
    \label{eq:appendix_posterior_raw_expression}
\end{align}
Grouping the fraction terms and exponential terms together and simplifying them, the RHS of Equation \eqref{eq:appendix_posterior_raw_expression} becomes
\begin{align}
    & \frac{1}{\sqrt{2\pi}\sqrt{\frac{\sigma^2_\epsilon\sigma^2_X}{\sigma^2_X + \sigma^2_\epsilon}}}
    \exp\Bigg(-\frac{1}{2}\Bigg[ \nonumber\\[-0.5em]
    & \quad \frac{\left(y - (x + \mu_\epsilon)\right)^2}{\sigma^2_\epsilon} + \frac{\left(x - \mu_X\right)^2}{\sigma^2_X} -
    \frac{\left(y - (\mu_X + \mu_\epsilon)\right)^2}{\sigma^2_X + \sigma^2_\epsilon}
    \Bigg]\Bigg)
    \label{eq:appendix_posterior_grouped_simplified}.
\end{align}
We then make $x$ the principal term in the squared expressions. Note we can rewrite $(y - (x + \mu_\epsilon))^2$ as $(x - (y - \mu_\epsilon))^2$ and $(y - (\mu_X + \mu_\epsilon))^2$ as $(\mu_X - (y - \mu_\epsilon))^2$ likewise. Expression~\eqref{eq:appendix_posterior_grouped_simplified} can then be written as
\begin{align}
    & \frac{1}{\sqrt{2\pi}\sqrt{\frac{\sigma^2_\epsilon\sigma^2_X}{\sigma^2_X + \sigma^2_\epsilon}}}
    \exp\Bigg(-\frac{1}{2}\Bigg[ \label{eq:appendix_posterior_grouped_simplified_x}\\
    & \quad \frac{\left(x - (y - \mu_\epsilon)\right)^2}{\sigma^2_\epsilon} + \frac{\left(x - \mu_X\right)^2}{\sigma^2_X} - 
    \frac{\left(\mu_X - (y - \mu_\epsilon))\right)^2}{\sigma^2_X + \sigma^2_\epsilon}
    \Bigg]\Bigg) .
    \nonumber
\end{align}

We now attempt to complete the square. The expression within the exponent can be grouped to the $x^2$, $x$, and constant terms, giving:
\begin{align}
    & \frac{1}{\sqrt{2\pi}\sqrt{\frac{\sigma^2_\epsilon\sigma^2_X}{\sigma^2_X + \sigma^2_\epsilon}}}
    \exp\bigg(-\frac{1}{2}\bigg[ \nonumber\\
    & \qquad x^2\left[\frac{1}{\sigma^2_\epsilon} \!+\! \frac{1}{\sigma^2_X} \right] - 
    2x\left[\frac{y-\mu_\epsilon}{\sigma^2_\epsilon} \!+\! 
            \frac{\mu_X}{\sigma^2_X}\right] + \nonumber\\
    & \qquad \frac{(y - \mu_\epsilon)^2}{\sigma^2_\epsilon} +
          \frac{\mu^2_X}{\sigma^2_X} -
          \frac{(\mu_X - (y - \mu_\epsilon))^2}{\sigma^2_X + \sigma^2_\epsilon}
    \bigg]\bigg).
    \label{eq:appendix_posterior_by_term}
\end{align}
Note we can rewrite the coefficient of each of the three terms with $\frac{\sigma^2_\epsilon \sigma^2_X}{\sigma^2_X + \sigma^2_\epsilon}$, the intended variance of the posterior distribution, as the denominator. The coefficient for the $x^2$ and $x$ terms are straightforward:
\begin{align}
    \frac{1}{\sigma^2_\epsilon} + \frac{1}{\sigma^2_X} = & 
    \frac{\sigma^2_X + \sigma^2_\epsilon}{\sigma^2_\epsilon \sigma^2_X} = 
    \frac{1}{\frac{\sigma^2_\epsilon \sigma^2_X}{\sigma^2_X + \sigma^2_\epsilon}} \;,
    \label{eq:appendix_posterior_x_sq_term}\\
    \frac{y-\mu_\epsilon}{\sigma^2_\epsilon} + 
            \frac{\mu_X}{\sigma^2_X} = & 
    \frac{\sigma^2_X (y - \mu_\epsilon) + \sigma^2_\epsilon(\mu_X)}{\sigma^2_\epsilon \sigma^2_X} \nonumber\\
    = & 
    \frac{\frac{\sigma^2_X}{\sigma^2_X + \sigma^2_\epsilon} (y - \mu_\epsilon) + \frac{\sigma^2_\epsilon}{\sigma^2_X + \sigma^2_\epsilon}(\mu_X)}{\frac{\sigma^2_\epsilon \sigma^2_X}{\sigma^2_X + \sigma^2_\epsilon}} \;.
    \label{eq:appendix_posterior_x_term}
\end{align}

To complete the square in Expression \eqref{eq:appendix_posterior_by_term} properly, we expect to see the numerator for the constant term to be the square of the numerator in the RHS of Expression \eqref{eq:appendix_posterior_x_term}. We first transform the constant term to one that contains the desired denominator:
\begin{align}
    & \frac{(y - \mu_\epsilon)^2}{\sigma^2_\epsilon} +
          \frac{\mu^2_X}{\sigma^2_X} -
          \frac{(\mu_X - (y - \mu_\epsilon))^2}{\sigma^2_X + \sigma^2_\epsilon} \nonumber\\[0.4em]
    = \, & \frac{\sigma^2_X (y - \mu_\epsilon)^2 + \sigma^2_\epsilon \mu^2_X - \frac{\sigma^2_\epsilon \sigma^2_X}{\sigma^2_X + \sigma^2_\epsilon}(\mu_X - (y - \mu_\epsilon))^2}{\sigma^2_\epsilon \sigma^2_X} \nonumber\\[0.4em]
    = \, & \frac{\frac{\sigma^2_X}{\sigma^2_X + \sigma^2_\epsilon} (y - \mu_\epsilon)^2 + \frac{\sigma^2_\epsilon}{\sigma^2_X + \sigma^2_\epsilon} \mu^2_X - \frac{\sigma^2_\epsilon \sigma^2_X}{(\sigma^2_X + \sigma^2_\epsilon)^2}(\mu_X \!-\! (y \!-\! \mu_\epsilon))^2}{\frac{\sigma^2_\epsilon \sigma^2_X}{\sigma^2_X + \sigma^2_\epsilon}} .
    \label{eq:appendix_posterior_constant_1}
\end{align}
This is followed by extracting the $(y - \mu_\epsilon)^2$ and $\mu^2_X$ terms from the numerator of Expression \eqref{eq:appendix_posterior_constant_1}, giving
\begin{align}
    \frac{1}{\frac{\sigma^2_\epsilon \sigma^2_X}{\sigma^2_X + \sigma^2_\epsilon}}
    \Bigg[ & 
    (y - \mu_\epsilon)^2 \left[\frac{\sigma^2_X}{\sigma^2_X + \sigma^2_\epsilon} - \frac{\sigma^2_\epsilon \sigma^2_X}{(\sigma^2_X + \sigma^2_\epsilon)^2}\right] + \nonumber\\
    & \mu^2_X \left[\frac{\sigma^2_\epsilon}{\sigma^2_X + \sigma^2_\epsilon} - \frac{\sigma^2_\epsilon \sigma^2_X}{(\sigma^2_X + \sigma^2_\epsilon)^2}\right] + \nonumber\\
    & 2\mu_X (y - \mu_\epsilon)\frac{\sigma^2_\epsilon \sigma^2_X}{(\sigma^2_X + \sigma^2_\epsilon)^2} \Bigg].
    \label{eq:appendix_posterior_constant_2}
\end{align}
By observing $\frac{\sigma^2_X}{\sigma^2_X + \sigma^2_\epsilon} - \frac{\sigma^2_\epsilon \sigma^2_X}{(\sigma^2_X + \sigma^2_\epsilon)^2} = \left(\frac{\sigma^2_X}{\sigma^2_X + \sigma^2_\epsilon}\right)^2$ and likewise for the $\sigma^2_\epsilon$ case, as well as expanding the last numerator term of Expression \eqref{eq:appendix_posterior_constant_2}, the constant term can be written as
\begin{align}
    \frac{1}{\frac{\sigma^2_\epsilon \sigma^2_X}{\sigma^2_X + \sigma^2_\epsilon}}
    \Bigg[ & 
    (y - \mu_\epsilon)^2 \left(\frac{\sigma^2_X}{\sigma^2_X + \sigma^2_\epsilon}\right)^2 + 
    \mu^2_X \left(\frac{\sigma^2_\epsilon}{\sigma^2_X + \sigma^2_\epsilon}\right)^2 + \nonumber\\
    & 2(y - \mu_\epsilon)\frac{\sigma^2_X}{\sigma^2_X + \sigma^2_\epsilon}\,\mu_X \frac{\sigma^2_\epsilon}{\sigma^2_X + \sigma^2_\epsilon} \Bigg],
    \label{eq:appendix_posterior_constant_3}
\end{align}
which the numerator is simply the expected result expanded:
\begin{align}
    & \frac{(y - \mu_\epsilon)^2}{\sigma^2_\epsilon} +
          \frac{\mu^2_X}{\sigma^2_X} -
          \frac{(\mu_X - (y - \mu_\epsilon))^2}{\sigma^2_X + \sigma^2_\epsilon} \nonumber\\
    = &
    \frac{\left(\frac{\sigma^2_X}{\sigma^2_X + \sigma^2_\epsilon} (y - \mu_\epsilon) + \frac{\sigma^2_\epsilon}{\sigma^2_X + \sigma^2_\epsilon}(\mu_X)\right)^2}{\frac{\sigma^2_\epsilon \sigma^2_X}{\sigma^2_X + \sigma^2_\epsilon}} \;.
    \label{eq:appendix_posterior_constant_4}
\end{align}

Substituting Equations \eqref{eq:appendix_posterior_x_sq_term}, \eqref{eq:appendix_posterior_x_term}, and \eqref{eq:appendix_posterior_constant_4} into Expression~\eqref{eq:appendix_posterior_by_term}, we arrive at the PDF of the posterior distribution:
\begin{align}
    & f_{X_n|Y_n} (x\,|\,y) \nonumber\\
    = \,&
    \frac{1}{\sqrt{2\pi}\!\sqrt{\sigma^2_{X_n|Y_n}}}
    \exp\left(\!-\frac{1}{2}\left[
    \frac{x^2 \!-\! 2x (\mu_{X_n | Y_n}) \!+\! (\mu_{X_n | Y_n})^2}{\sigma^2_{X_n|Y_n}}
    \right]\right) \nonumber\\
    = \,& \frac{1}{\sqrt{2\pi}\sqrt{\sigma^2_{X_n|Y_n}}}
    \exp\left(-\frac{1}{2}\left[
    \frac{(x - \mu_{X_n | Y_n})^2}{\sigma^2_{X_n|Y_n}}
    \right]\right),
\end{align}
where $\mu_{X_n | Y_n}$ and $\sigma^2_{X_n|Y_n}$ are that defined in Equations~\eqref{eq:appendix_posterior_mu} and~\eqref{eq:appendix_posterior_sigma_sq} respectively. This is clearly the PDF of a normal distribution.

\subsection{Covariance of a conditioned normal r.v.}
\label{sec:2d}

In this section we state the standard Bayesian inference result in the multi-dimensional case. We then use the generalized result to show that if the $X_n$s and $Y_n$s are set up as above, any $X_i$ and $X_j$ ($i\neq j$) will still be uncorrelated even when the values of the corresponding $Y_i$ and $Y_j$ are known. This result will be used when we derive Equation~\eqref{eq:cov_xir_xis_final} in full in Section~\ref{sec:appendix_var_D}.\\

\noindent\textbf{The general result:}

Eaton \cite{eaton83multivariate} has provided the following result regarding the conditional distribution for a multivariate normal distribution.
Let $\bm{x}$ be a $n$-dimensional multivariate normal random variable. If we partition $\bm{x}$ into two components of dimension $q$ and $n-q$, and its mean and covariance matrix accordingly such that
\begin{align}
    \bm{x} = \left(\!\!\begin{array}{c} \bm{x}_1 \\ \bm{x}_2 \end{array} \!\!\right) 
    \sim \mathcal{N}
    \bigg( \bm{\mu} = \left(\!\!\begin{array}{c} \bm{\mu}_1 \\ \bm{\mu}_2 \end{array}\!\!\right) \,,\,
    \bm{\Sigma} = \left(\!\!\begin{array}{cc} \bm{\Sigma}_{11} & \bm{\Sigma}_{12} \\ \bm{\Sigma}_{21} & \bm{\Sigma}_{22} \end{array} \!\!\right)\bigg),
\end{align}
where $\bm{x}_1, \bm{\mu}_1 \in \mathbb{R}^q$, $\bm{x}_1, \bm{\mu}_1 \in \mathbb{R}^{n-q}$, and $\bm{\Sigma}_{11}$, $\bm{\Sigma}_{12}$, $\bm{\Sigma}_{21}$, and $\bm{\Sigma}_{22}$ have sizes $q\times q$, $q \times (n-q)$, $(n-q) \times q$, and $(n-q) \times (n-q)$ respectively, then the distribution of $\bm{x}_1$ conditional on $\bm{x}_2 = \bm{a}$ is also a multivariate normal with:
\begin{align}
    & (\bm{x}_1 | \bm{x}_2 = \bm{a}) \nonumber\\
    \sim \,&  \mathcal{N}\big(
    \bm{\mu}_1 + \bm{\Sigma}_{12} \bm{\Sigma}^{-1}_{22} (\bm{a} - \bm{\mu}_2) \,,\,
    \bm{\Sigma}_{11} - \bm{\Sigma}_{12} \bm{\Sigma}^{-1}_{22} \bm{\Sigma}_{21}\big) .
\end{align}\\

\noindent\textbf{Specializing the result to our setup:}

To show the uncorrelatedness of $X_n$s conditional on corresponding $Y_n$s, we revisit the setup in Section \ref{sec:1d} and focus on the $i^{\textrm{th}}$- and $j^{\textrm{th}}$-indexed random variables. This allows us to construct a four-dimensional vector ${\bm{x} = (X_i, X_j, Y_i, Y_j)^T}$ and partition them into two two-dimensional components, so that the general result can be applied.

The setup specifies that ${X_n \overset{\textrm{i.i.d.}}{\sim} \mathcal{N}(\mu_X, \sigma^2_X)}$, ${\epsilon_n \overset{\textrm{i.i.d.}}{\sim} \mathcal{N}(\mu_\epsilon, \sigma^2_\epsilon)}$, and ${X_i \perp \epsilon_j \:\forall\, i, j}$. For the $i^{\textrm{th}}$- and $j^{\textrm{th}}$-indexed random variables we have
\begin{align}
    X_i \sim \mathcal{N}(\mu_X, \sigma^2_X) & \,,\, X_j \sim \mathcal{N}(\mu_X, \sigma^2_X),  \label{eq:appendix_X_i_distribution}\\
    \epsilon_i \sim \mathcal{N}(\mu_\epsilon, \sigma^2_\epsilon) & \,,\, \epsilon_j \sim \mathcal{N}(\mu_\epsilon, \sigma^2_\epsilon), \label{eq:appendix_epsilon_i_distribution}
\end{align}
where all the four random variables are independent from each other.
The setup also specifies that $Y_n = X_n + \epsilon_n$, this yields
\begin{align}
    Y_i \sim \mathcal{N}(\mu_X \!+\! \mu_\epsilon,\, \sigma^2_X \!+\! \sigma^2_\epsilon),\, \textrm{and} \;
    Y_j \sim \mathcal{N}(\mu_X \!+\! \mu_\epsilon,\, \sigma^2_X \!+\! \sigma^2_\epsilon),
    \label{eq:appendix_Y_i_distribution}
\end{align}
where $X_i \perp Y_j$ $\forall i \neq j$.

We then construct the mean vector and covariance matrix required by the general result. Expressions \eqref{eq:appendix_X_i_distribution} and \eqref{eq:appendix_Y_i_distribution} provided the information required to complete the entire mean vector and most of the covariance matrix. We only need to obtain $\textrm{Cov}(X_i, Y_i)$ (and $\textrm{Cov}(X_j, Y_j)$, which amounts to the same quantity). By definition of covariance:
\begin{align}
    \textrm{Cov}(X_i, Y_i) = \mathbb{E}(X_iY_i) - \mathbb{E}(X_i)\mathbb{E}(Y_i).
    \label{eq:appendix_cov_xi_yi}
\end{align}
The first term is calculated by observing $Y_i = X_i + \epsilon_i$, $X_i \perp \epsilon_i$, and standard identities between expectations and variances:
\begin{align}
    \mathbb{E}(X_iY_i)
    =\, &  \mathbb{E}(X_i (X_i + \epsilon_i)) 
    = \mathbb{E}(X_i^2) + \mathbb{E}(X_i\epsilon_i) \nonumber \\
    =\, & \mathbb{E}^2(X_i) + \textrm{Var}(X_i) + \mathbb{E}(X_i)\mathbb{E}(\epsilon_i) 
    \label{eq:appendix_e_xi_yi}.
\end{align}
Using Expressions \eqref{eq:appendix_X_i_distribution} and \eqref{eq:appendix_epsilon_i_distribution} when substituting the expectations and variances, we have from the above
\begin{align}
    & \textrm{Cov}(X_i, Y_i) \nonumber\\
    = & \, \mathbb{E}^2(X_i) + \textrm{Var}(X_i) + \mathbb{E}(X_i)\mathbb{E}(\epsilon_i) - \mathbb{E}(X_i)\mathbb{E}(Y_i) \nonumber\\
    = & \, \mu_X^2 + \sigma^2_X + \mu_X\mu_\epsilon - \mu_X(\mu_X + \mu_\epsilon) = \sigma^2_X .
\end{align}
This yields the mean vector and covariance matrix for ${\bm{x} = (X_i, X_j, Y_i, Y_j)^T}$ as
\begin{align}
    \bm{\mu} = \left(\!\!\begin{array}{c} \mu_X \\ \mu_X \\ \mu_X + \mu_\epsilon \\ \mu_X + \mu_\epsilon \end{array}\!\!\right),\,
    \bm{\Sigma} = \left(\!\!\begin{array}{cccc} \sigma^2_X\! & 0\!\! & \sigma^2_X\!\! & 0 \\ 0\! & \sigma^2_X\!\! & 0\!\! & \sigma^2_X \\ \sigma^2_X\! & 0 & \sigma^2_X + \sigma^2_\epsilon\!\! & 0 \\ 0\! & \sigma^2_X\!\! & 0\!\! & \sigma^2_X + \sigma^2_\epsilon \end{array}\!\!\right).
\end{align}

Partitioning $\bm{x}$ into two components $(X_i, X_j)^T$ and $(Y_i, Y_j)^T$, and using the general result above we then have:
\begin{align}
    \left(\begin{array}{c} X_i \\ X_j \end{array}\right) \Bigg| \left(\begin{array}{c} Y_i \\ Y_j \end{array}\right) = \left(\begin{array}{c} y_i \\ y_j \end{array}\right) 
    \sim \mathcal{N} \left(\bm{\mu}', \bm{\Sigma}'\right)
\end{align}
where
\begin{fleqn}
\begin{align}
    \bm{\mu}' = & \binom{\mu_X}{\mu_X}+ 
    \left(\begin{array}{cc} \sigma^2_X & 0 \\ 0 & \sigma^2_X \end{array}\right) \cdot \nonumber\\
    & \quad \left(\!\!\begin{array}{cc} \sigma^2_X + \sigma^2_\epsilon & 0 \\ 0 & \sigma^2_X + \sigma^2_\epsilon \end{array}\!\!\right)^{-1} \left[\binom{y_i}{y_j} - \binom{\mu_X + \mu_\epsilon}{\mu_X + \mu_\epsilon} \right], \nonumber
\end{align}
\begin{align}
    \bm{\Sigma}' = & \left(\begin{array}{cc} \sigma^2_X & 0 \\ 0 & \sigma^2_X \end{array}\right) - \left(\begin{array}{cc} \sigma^2_X & 0 \\ 0 & \sigma^2_X \end{array}\right) \cdot \nonumber\\
    & \quad \left(\!\!\begin{array}{cc} \sigma^2_X + \sigma^2_\epsilon & 0 \\ 0 & \sigma^2_X + \sigma^2_\epsilon \end{array}\!\!\right)^{-1} \left(\begin{array}{cc} \sigma^2_X & 0 \\ 0 & \sigma^2_X \end{array}\right) .
\end{align}
\end{fleqn}
Simplifying the expression by standard matrix operations we arrive at:
\begin{align}
    \bm{\mu}' = & \left(\begin{array}{c} \frac{\sigma^2_\epsilon}{\sigma^2_X + \sigma^2_\epsilon} \mu_X + \frac{\sigma^2_X}{\sigma^2_X + \sigma^2_\epsilon} (y_i - \mu_\epsilon) \\ \frac{\sigma^2_\epsilon}{\sigma^2_X + \sigma^2_\epsilon} \mu_X + \frac{\sigma^2_X}{\sigma^2_X + \sigma^2_\epsilon} (y_j - \mu_\epsilon) \end{array}\right), \\
    \bm{\Sigma}' = & \left(\begin{array}{cc} \frac{\sigma^2_\epsilon \sigma^2_X}{\sigma^2_X + \sigma^2_\epsilon } & 0 \\ 0 & \frac{\sigma^2_\epsilon \sigma^2_X}{\sigma^2_X + \sigma^2_\epsilon } \end{array}\right) .
\end{align}
Note the mean and variance of $X_i$ and $X_j$ is that derived in Section \ref{sec:1d}. Furthermore, the covariance between $X_i$ and $X_j$ given $Y_i$ and $Y_j$ is zero, as claimed at the beginning of this subsection.

\subsection{Calculating the Expectation}
\label{sec:appendix_gauss_mean}

The results above are necessary to derive the quantities presented in Section~\ref{sec:normal_normal_model} of the paper. Here we go through the derivation in greater detail, focusing more on the algebraic manipulations and being light on the commentaries found in the main paper. Readers are encouraged to read Section~\ref{sec:mathematical_formulation} of the paper to familiarize themselves with the notations and terminologies used below before proceeding. 

We are interested in deriving the expected value of $D$, the value gained when the estimation noise is reduced. To do so we require the expected values of, in order:
\begin{enumerate}
    \item $Y_{(r)}$ - the \emph{estimated} value of the $r^{\textrm{th}}$ proposition, ranked in increasing estimated value;
    \item $X_{\mathcal{I}(r)}$ - the \emph{true} value of the $r^{\textrm{th}}$ proposition, ranked by increasing estimated value
    \item $V$ - the mean of the \emph{true} value for the $M$ most valuable propositions, ranked by their estimated values.
\end{enumerate}  

To obtain the expected value for $Y_{(r)}$, we begin by observing that the $Y_n$ are normally distributed. Using the standard properties of the normal distribution:
\begin{align}
    Y_n = X_n + \epsilon_n & \overset{\textrm{i.i.d.}}{\sim} \mathcal{N}(\mu_X + \mu_\epsilon,\, \sigma^2_X + \sigma^2_\epsilon).
    \label{eq:appendix_yn_distribution}
\end{align}
This is followed by applying a result by Blom \cite{blom58statisticalestimates}, which states that the expected value for normal order statistics $Y_{(r)}$ can be closely approximated as:
\begin{align}
    \mathbb{E}(Y_{(r)}) \approx 
    \mu_X + \mu_\epsilon + 
    \sqrt{\sigma^2_X + \sigma^2_\epsilon}\;\Phi^{-1}\left(\frac{r - \alpha}{N - 2\alpha + 1}\right),
    \label{eq:appendix_observed_val_normal_order_stat_approx}
\end{align}
where $\Phi^{-1}$ denotes the quantile function of a standard normal distribution, and $\alpha$ is a constant.\footnote{While there are a number of literature on the best value of~$\alpha$, the exact value of this constant does not play a huge role as long as $\alpha \approx 0.4$~\cite{harter61expectedvalues}.}

The expected value of $X_{\mathcal{I}(r)}$ is obtained as follows. We first recall two standard results in Bayesian inference, the first being the conditional probability distribution for any $Y_n$ given $X_n$ is normally distributed:
\begin{align}
Y_n | (X_n = x) & \; \sim \; \mathcal{N}(x + \mu_\epsilon, \sigma^2_\epsilon).
\end{align}
The second result states that the posterior distribution of $X_n$ once $Y_n$ is observed is also normally distributed, with mean $\mu_{X_n | Y_n}$ and variance $\sigma^2_{X_n | Y_n}$ given by (see Section~\ref{sec:1d}):
\begin{align}
    \mu_{X_n | \left (Y_n=y \right )} & = \frac{\sigma^2_X}{\sigma^2_X + \sigma^2_\epsilon} (y - \mu_\epsilon) + \frac{\sigma^2_\epsilon}{\sigma^2_X + \sigma^2_\epsilon} \mu_X 
    \label{eq:appendix_true_val_posterior_mean},
    \\
    \sigma^2_{X_n | Y_n} & = \frac{\sigma^2_\epsilon \sigma^2_X}{\sigma^2_X + \sigma^2_\epsilon} \,.
    \label{eq:appendix_true_val_posterior_variance}
\end{align}
We then apply the law of iterated expectations to obtain 
\begin{align}
    \mathbb{E}(X_{\mathcal{I}(r)}) = \mathbb{E}_{Y_{(r)}}\left(\mathbb{E}(X_{\mathcal{I}(r)} \,|\, Y_{(r)})\right).
    \label{eq:appendix_true_val_law_iter_exp}
\end{align}
Noting $Y_{\mathcal{I}(r)}$ is equivalent to $Y_{(r)}$ as the propositions are ranked by their estimated values, we substitute Equation~\eqref{eq:appendix_true_val_posterior_mean} into Equation~\eqref{eq:appendix_true_val_law_iter_exp}, and move the constant terms out of the outer expectation to get
\begin{align}
    \mathbb{E}(X_{\mathcal{I}(r)}) & = 
    \mathbb{E}_{Y_{(r)}}\left(\frac{\sigma^2_X}{\sigma^2_X + \sigma^2_\epsilon} (y - \mu_\epsilon) + \frac{\sigma^2_\epsilon}{\sigma^2_X + \sigma^2_\epsilon} \mu_X \right) \nonumber\\
    & = \frac{\sigma^2_X}{\sigma^2_X + \sigma^2_\epsilon} \mathbb{E}_{Y_{(r)}}(y) - 
    \frac{\sigma^2_X}{\sigma^2_X + \sigma^2_\epsilon} \mu_\epsilon+
    \frac{\sigma^2_\epsilon}{\sigma^2_X + \sigma^2_\epsilon} \mu_X.
    \label{eq:appendix_true_val_law_iter_exp_expanded}
\end{align}
Further substituting Equation~\eqref{eq:appendix_observed_val_normal_order_stat_approx} into Equation~\eqref{eq:appendix_true_val_law_iter_exp_expanded} and simplifying the equation yields
\begin{align}
    & \mathbb{E}(X_{\mathcal{I}(r)}) \nonumber \\ \approx \, &
    \frac{\sigma^2_X}{\sigma^2_X + \sigma^2_\epsilon} \left(\mu_X + \mu_\epsilon + 
    \sqrt{\sigma^2_X + \sigma^2_\epsilon}\;\Phi^{-1}\left(\frac{r - \alpha}{N - 2\alpha + 1}\right)\right) \nonumber\\ & - 
    \frac{\sigma^2_X}{\sigma^2_X + \sigma^2_\epsilon} \mu_\epsilon+
    \frac{\sigma^2_\epsilon}{\sigma^2_X + \sigma^2_\epsilon}\mu_X.
    \label{eq:appendix_true_val_law_iter_exp_substituted}
\end{align}
The $\mu_\epsilon$ terms cancel, and the $\mu_X$ terms sum to $\mu_X$ leading to
\begin{align}
    \mathbb{E}(X_{\mathcal{I}(r)}) & \approx 
    \mu_X + 
    \frac{\sigma^2_X  \sqrt{\sigma^2_X + \sigma^2_\epsilon}}{\sigma^2_X + \sigma^2_\epsilon}  
   \;\Phi^{-1}\left(\frac{r - \alpha}{N - 2\alpha + 1}\right) \nonumber\\
   & = \mu_X + 
    \frac{\sigma^2_X}{\sqrt{\sigma^2_X + \sigma^2_\epsilon}}
   \;\Phi^{-1}\left(\frac{r - \alpha}{N - 2\alpha + 1}\right).
    \label{eq:appendix_true_val}
\end{align}

Equation~\eqref{eq:appendix_true_val} shows that decreasing the estimation noise~$\sigma^2_\epsilon$ will lead to an increase in $\mathbb{E}(X_{\mathcal{I}(r)})$ for any $r > \frac{N+1}{2}$. It follows that the mean true value of the top $M$ propositions, selected according to their estimated value, will increase with the presence of a lower estimation noise. We show this by applying the expectation function to $V$ defined in Equation~\eqref{eq:mean_true_value}:
\begin{align}
    \mathbb{E}(V) = \frac{1}{M}\big(&\mathbb{E}(X_{\mathcal{I}(N-M+1)}) + \mathbb{E}(X_{\mathcal{I}(N-M+2)}) + \nonumber\\ & \cdots + \mathbb{E}(X_{\mathcal{I}(N)})\big).
\end{align}
Substituting Equation~\eqref{eq:appendix_true_val} into the above gives
\begin{align}
   & \mathbb{E}(V) \nonumber\\
   \approx \, & \frac{1}{M}\bigg(\mu_X + 
    \frac{\sigma^2_X}{\sqrt{\sigma^2_X + \sigma^2_\epsilon}}
   \;\Phi^{-1}\left(\frac{N-M+1 - \alpha}{N - 2\alpha + 1}\right) + \nonumber\\
    & \qquad \mu_X + 
    \frac{\sigma^2_X}{\sqrt{\sigma^2_X + \sigma^2_\epsilon}}
   \;\Phi^{-1}\left(\frac{N-M+2 - \alpha}{N - 2\alpha + 1}\right) + \cdots + \nonumber\\
   & \qquad \mu_X + 
    \frac{\sigma^2_X}{\sqrt{\sigma^2_X + \sigma^2_\epsilon}}
   \;\Phi^{-1}\left(\frac{N - \alpha}{N - 2\alpha + 1}\right)\bigg).
   \label{eq:appendix_expected_mean_true_value}
\end{align}
Observing there are $M$ copies of $\mu_X$, and the $\Phi^{-1}$ terms can be written as a summation, we arrive at
\begin{align}
    \mathbb{E}(V) \approx \mu_X + \frac{\sigma^2_X}{\sqrt{\sigma^2_X + \sigma^2_\epsilon}} \frac{1}{M} \sum_{r=N-M+1}^{N} \Phi^{-1}\left(\frac{r-\alpha}{N - 2\alpha + 1}\right).
    \label{eq:appendix_expected_mean_true_value_simplified}
\end{align}

We finally consider the improvement when we reduce the estimation noise from $\sigma^2_\epsilon = \sigma^2_1$ to $\sigma^2_2$:
\begin{align}
    & \mathbb{E}(D) = \mathbb{E}(V|_{\sigma^2_\epsilon = \sigma^2_2}) - \mathbb{E}(V|_{\sigma^2_\epsilon = \sigma^2_1}) \nonumber\\
    \approx & \left(\frac{\sigma^2_X}{\sqrt{\sigma^2_X + \sigma^2_2}} - \frac{\sigma^2_X}{\sqrt{\sigma^2_X + \sigma^2_1}} \right) \times \nonumber\\
    & \quad\frac{1}{M} \sum_{r=N-M+1}^{N} \Phi^{-1}\left(\frac{r-\alpha}{N - 2\alpha + 1}\right).
    \label{eq:appendix_expected_mean_true_value_diff}
\end{align}
If we assume $\mu_X = 0$ (i.e. the true value of the propositions are centred around zero), then the relative gain is entirely dependent on $\sigma^2_X$, $\sigma^2_1$, $\sigma^2_2$:
\begin{align}
    & \frac{\mathbb{E}(D|_{\mu_X = 0})}{\mathbb{E}(V|_{\sigma^2_\epsilon = \sigma^2_1,\, \mu_X = 0})} 
    = \, \frac{\sqrt{\sigma^2_X + \sigma^2_1}}{\sqrt{\sigma^2_X + \sigma^2_2}} - 1 \,.
\label{eq:appendix_gauss_res}
\end{align}


\subsection{Calculating the Variance}
\label{sec:appendix_var_D}
Having derived the expected value in~Equations~\eqref{eq:appendix_expected_mean_true_value_diff} and~\eqref{eq:appendix_gauss_res}, in this section we address
the variance of $D$. Deriving the variance is similar to deriving the expectation --- one has to obtain the variances for (in order) $Y_{(r)}$, $X_{\mathcal{I}(r)}$, and $V$.
For the variance of $Y_{(r)}$, we apply a result from David and Johnson~\cite{david54statisticaltreatment}, which states that given $Y_n$ as defined in Expression~\eqref{eq:appendix_yn_distribution}, $\textrm{Var} \left ( Y_{(r)} \right )$ can be approximated as:
\begin{align}
    \textrm{Var}(Y_{(r)}) \approx
    \frac{r(N-r+1)}{(N+1)^2 (N+2)}
    \frac{\sigma^2_X + \sigma^2_\epsilon}{\left(\phi\left(\Phi^{-1}\left(\frac{r}{N+1}\right)\right)\right)^2} \,,
    \label{eq:appendix_observed_val_normal_order_stat_variance_approx}
\end{align}
where $\phi$ is the probability density function, and $\Phi^{-1}$ is the quantile function of a standard normal distribution.

The variance for $X_{\mathcal{I}(r)}$ is then obtained using the law of total variance:
\begin{align}
    & \textrm{Var}(X_{\mathcal{I}(r)}) \nonumber \\ 
    =\, & 
    \mathbb{E}_{Y_{(r)}}\left(\textrm{Var}(X_{\mathcal{I}(r)}|Y_{(r)})\right) +
    \textrm{Var}_{Y_{(r)}}\left(\mathbb{E}(X_{\mathcal{I}(r)}|Y_{(r)})\right) .
    \label{eq:appendix_true_val_law_total_var}
\end{align}
Recognizing $Y_{(r)}$ and $Y_{\mathcal{I}(r)}$ are equivalent, we substitute the conditional expectations stated in Equations~\eqref{eq:appendix_true_val_posterior_mean} and~\eqref{eq:appendix_true_val_posterior_variance} into Equation~\eqref{eq:appendix_true_val_law_total_var}, and move the constant terms out of the outer expectation / variance to get
\begin{align}
    \textrm{Var}(X_{\mathcal{I}(r)}) 
    & = 
    \mathbb{E}_{Y_{(r)}}\left(\frac{\sigma^2_\epsilon \sigma^2_X}{\sigma^2_X + \sigma^2_\epsilon}\right) + \nonumber \\
    & \quad\; \textrm{Var}_{Y_{(r)}}\left(\frac{\sigma^2_X}{\sigma^2_X + \sigma^2_\epsilon} (y - \mu_\epsilon) + \frac{\sigma^2_\epsilon}{\sigma^2_X + \sigma^2_\epsilon} \mu_X \right) \nonumber\\
    & = \frac{\sigma^2_\epsilon \sigma^2_X}{\sigma^2_X + \sigma^2_\epsilon} + 
    \left(\frac{\sigma^2_X}{\sigma^2_X + \sigma^2_\epsilon}\right)^2 \textrm{Var}_{Y_{(r)}}(y) \;.
    \label{eq:appendix_true_val_law_total_var_expanded}
\end{align}
Further substituting Expression~\eqref{eq:appendix_observed_val_normal_order_stat_variance_approx} into Equation~\eqref{eq:appendix_true_val_law_total_var_expanded} and cancelling out equal terms, we have
\begin{align}
    & \textrm{Var}(X_{\mathcal{I}(r)}) \approx 
    \frac{\sigma^2_\epsilon \sigma^2_X}{\sigma^2_X + \sigma^2_\epsilon} + \nonumber \\
    & \quad \frac{\sigma^4_X}{\sigma^2_X + \sigma^2_\epsilon} \frac{r(N-r+1)}{(N+1)^2 (N+2)}
    \frac{1}{\left(\phi\left(\Phi^{-1}(\frac{r}{N+1})\right)\right)^2} \,.
    \label{eq:appendix_var_xir}
\end{align}

Before we derive the variance of $V$, we require the covariance between pairs of~$Y_{(\cdot)}$s and~$X_{\mathcal{I}(\cdot)}$s. This is necessary as the terms of $V$ (see Equation \eqref{eq:mean_true_value}), being the result of removing noise from successive order statistics, are highly correlated.

David and Nagaraja \cite{david04orderstatistics} have provided a formula to estimate the covariance between~$Y_{(r)}$ and~$Y_{(s)}$ for any $r,s\leq N$ , first presented by David and Johnson \cite{david54statisticaltreatment}:
\begin{align}
    & \textrm{Cov}(Y_{(r)}, Y_{(s)}) \nonumber\\ \approx \,&
    \frac{r(N-s+1)}{(N+1)^2(N+2)}
    \frac{\sigma^2_X + \sigma^2_\epsilon}{\phi\left(\Phi^{-1}(\frac{r}{N+1})\right)\,\phi\left(\Phi^{-1}(\frac{s}{N+1})\right)} \,.
    \label{eq:appendix_cov_yr_ys}
\end{align}

To obtain the covariance between $X_{\mathcal{I}(r)}$ and $X_{\mathcal{I}(s)}$ for any $r, s \leq N$, we first observe the law of total covariance with multiple conditioning variables \cite{bowsher12identifyingsources} states that for $r \neq s$:
\begin{align}
    & \textrm{Cov}(X_{\mathcal{I}(r)}, X_{\mathcal{I}(s)})  \nonumber \\
    = \, &
    \mathbb{E}\left(\mathbb{E}(\textrm{Cov}(X_{\mathcal{I}(r)}, X_{\mathcal{I}(s)} \,|\, Y_{(r)}, Y_{(s)}) \,|\, Y_{(r)})\right) + \nonumber \\
    & \mathbb{E}\big(\textrm{Cov}(\mathbb{E}(X_{\mathcal{I}(r)} \,|\, Y_{(r)}, Y_{(s)}), \nonumber \\
    & \qquad \quad \; \mathbb{E}(X_{\mathcal{I}(s)} \,|\, Y_{(r)}, Y_{(s)}) \,|\, Y_{(r)} )\big) \, + \nonumber \\
    & \textrm{Cov}\big(\mathbb{E}(\mathbb{E}(X_{\mathcal{I}(r)} \,|\, Y_{(r)}, Y_{(s)}) \,|\, Y_{(r)}), \nonumber\\ 
    & \qquad  \: \mathbb{E}(\mathbb{E}(X_{\mathcal{I}(s)} \,|\, Y_{(r)}, Y_{(s)}) \,|\, Y_{(r)})\big) \,.
    \label{eq:appendix_cov_xir_xis}
\end{align}
Noting that $Y_{(\cdot)}$ is equivalent to $Y_{\mathcal{I}(\cdot)}$ by definition, we observe the first term on the RHS of Equation \eqref{eq:appendix_cov_xir_xis} is zero. This follows from properties of a multivariate normal's conditional distributions --- $X_n$s are uncorrelated to each other given the corresponding $Y_n$s if $X_n$ themselves are uncorrelated (see Section~\ref{sec:2d}). For the second and third term, we note $X_n$ is independent of $Y_m$ for any $n \neq m$ and hence $\mathbb{E}(X_n | Y_n, Y_m) = \mathbb{E}(X_n | Y_n)$. This allows us to substitute Equation \eqref{eq:appendix_true_val_posterior_mean} into Equation \eqref{eq:appendix_cov_xir_xis} to get
\begin{align}
    & \textrm{Cov}(X_{\mathcal{I}(r)}, X_{\mathcal{I}(s)}) = 0 \,+ \nonumber \\
    & \quad \mathbb{E}\bigg(\textrm{Cov}\bigg(\frac{\sigma^2_X}{\sigma^2_X + \sigma^2_\epsilon} (Y_{(r)} - \mu_\epsilon) + \frac{\sigma^2_\epsilon}{\sigma^2_X + \sigma^2_\epsilon} \mu_X, \nonumber\\
    & \quad \qquad \qquad \frac{\sigma^2_X}{\sigma^2_X + \sigma^2_\epsilon} (Y_{(s)} - \mu_\epsilon) + \frac{\sigma^2_\epsilon}{\sigma^2_X + \sigma^2_\epsilon} \mu_X \,\bigg|\, Y_{(r)}\bigg)\bigg) + \nonumber \\
    & \quad \textrm{Cov}\bigg(\mathbb{E}\bigg(\frac{\sigma^2_X}{\sigma^2_X + \sigma^2_\epsilon} (Y_{(r)} - \mu_\epsilon) + \frac{\sigma^2_\epsilon}{\sigma^2_X + \sigma^2_\epsilon} \mu_X \,\bigg|\, Y_{(r)} \bigg), \nonumber\\
    & \quad \qquad \;\, \mathbb{E}\bigg(\frac{\sigma^2_X}{\sigma^2_X + \sigma^2_\epsilon} (Y_{(s)} - \mu_\epsilon) + \frac{\sigma^2_\epsilon}{\sigma^2_X + \sigma^2_\epsilon} \mu_X \,\bigg|\, Y_{(r)} \bigg)\bigg),
\end{align}
which can be simplified using standard properties of expectation and covariance functions to give
\begin{align}
    & \textrm{Cov}(X_{\mathcal{I}(r)}, X_{\mathcal{I}(s)}) \nonumber \\ 
    = \, & \mathbb{E}\left(\left(\frac{\sigma^2_X}{\sigma^2_X + \sigma^2_\epsilon}\right)^2 \textrm{Cov}\left(Y_{(r)}, Y_{(s)} \,|\, Y_{(r)}\right)\right) + \nonumber\\
    & \quad\left(\frac{\sigma^2_X}{\sigma^2_X + \sigma^2_\epsilon}\right)^2 \textrm{Cov}\left(Y_{(r)}, Y_{(s)}\right) .
    \label{eq:appendix_cov_xir_xis_simplified}
\end{align}
Clearly $\textrm{Cov}(A, B \,|\, A) = 0 \,\,\, \forall A, B$. Hence we substitute Equation \eqref{eq:appendix_cov_yr_ys} into Equation \eqref{eq:appendix_cov_xir_xis_simplified} and simplify the resultant expression to arrive at
\begin{align}
    & \textrm{Cov}(X_{\mathcal{I}(r)}, X_{\mathcal{I}(s)}) \approx \nonumber \\ 
    & \frac{\sigma^4_X}{\sigma^2_X + \sigma^2_\epsilon} \frac{r(N-s+1)}{(N+1)^2(N+2)}\frac{1}{\phi\left(\Phi^{-1}(\frac{r}{N+1})\right)\,\phi\left(\Phi^{-1}(\frac{s}{N+1})\right)} .
    \label{eq:appendix_cov_xir_xis_final}
\end{align}
Equation~\eqref{eq:appendix_cov_xir_xis_final} affirms the claim that the $X_{\mathcal{I}(\cdot)}$s are positively correlated. 
%
Now we can state the variance of $V$ and $D$. Applying the variance function to the defintion of $V$ (Equation~\eqref{eq:mean_true_value}) we get
\begin{align}
    & \textrm{Var}(V) \nonumber\\
    =\, &\frac{1}{M^2}\textrm{Var}\left(X_{\mathcal{I}(N-M+1)} + X_{\mathcal{I}(N-M+2)} + ... + X_{\mathcal{I}(N)}\right) \nonumber\\
    =\, & \frac{1}{M^2}\Bigg(\sum_{r=N-M+1}^{N} \textrm{Var}\left(X_{\mathcal{I}(r)}\right) \, + \nonumber\\
    & \qquad \;\: \sum_{r=N-M+1}^{N} \, \sum_{s=r+1}^{N} 2 \cdot \textrm{Cov}\left(X_{\mathcal{I}(r)}, X_{\mathcal{I}(s)}\right)\Bigg) ,
    \label{eq:appendix_var_V}
\end{align}
where $\textrm{Var}(X_{\mathcal{I}(r)})$ and $\textrm{Cov}(X_{\mathcal{I}(r)}, X_{\mathcal{I}(s)})$ are defined in Equations~\eqref{eq:appendix_var_xir} and \eqref{eq:appendix_cov_xir_xis_final}.

The variance of $D$ is thus:
\begin{align}
    \textrm{Var}(D) = & \textrm{Var}(V |_{\sigma^2_\epsilon = \sigma^2_2} - V |_{\sigma^2_\epsilon = \sigma^2_1}) \nonumber\\
    = & \textrm{Var}\left(V |_{\sigma^2_\epsilon = \sigma^2_2}\right) + \textrm{Var}\left(V |_{\sigma^2_\epsilon = \sigma^2_1}\right) \nonumber\\
    & \qquad - 2 \cdot \textrm{Cov}\left(V |_{\sigma^2_\epsilon = \sigma^2_2}, V |_{\sigma^2_\epsilon = \sigma^2_1}\right) .
    \label{eq:appendix_var_D}
\end{align}
The first two terms on the right hand side of Equation~\eqref{eq:appendix_var_D} are that defined in Equation~\eqref{eq:appendix_var_V} (we omit the expanded form for brevity), while the last term can be expanded as follow:
\begin{align}
   & \textrm{Cov}\left(V |_{\sigma^2_\epsilon = \sigma^2_2}, V |_{\sigma^2_\epsilon = \sigma^2_1}\right) \nonumber\\
   = \,&  \textrm{Cov}\Big(\frac{1}{M}\big(X_{\mathcal{I}(N-M+1)} \!+\! X_{\mathcal{I}(N-M+2)} \!+\! ... \!+\! 
   X_{\mathcal{I}(N)}\big) |_{\sigma^2_\epsilon = \sigma^2_2},  \nonumber \\ 
   & \qquad\; \frac{1}{M}\big(X_{\mathcal{I}(N-M+1)} \!+\! X_{\mathcal{I}(N-M+2)} \!+\! ... \!+\! X_{\mathcal{I}(N)}\big) |_{\sigma^2_\epsilon = \sigma^2_1}\Big) \nonumber\\
   = \,& \frac{1}{M^2}\sum_{r=N-M+1}^{N} \sum_{s=r}^{N} \, 2 \cdot \textrm{Cov}\left(X_{\mathcal{I}(r)} |_{\sigma^2_\epsilon = \sigma^2_2},\, X_{\mathcal{I}(s)} |_{\sigma^2_\epsilon = \sigma^2_1}\right) \,.
   \label{eq:appendix_cov_v_diff_noise}
\end{align}
Equation~\eqref{eq:appendix_cov_v_diff_noise} shows the covariance term in Equation~\eqref{eq:appendix_var_D} is positive as all its components are positive (cf. Equation~\eqref{eq:appendix_cov_xir_xis_final}, albeit with a different magnitude). Hence the variance terms in Equation~\eqref{eq:appendix_var_D} form an upper bound to the variance of~$D$:
\begin{align}
     \textrm{Var}(D) < 
     \textrm{Var}\left(V |_{\sigma^2_\epsilon = \sigma^2_2}\right) + \textrm{Var}\left(V |_{\sigma^2_\epsilon = \sigma^2_1}\right) .
\end{align}
In practice, the variance of $D$ is much lower than the bound, due to the $V$s being highly correlated. 


\section{Empirical Extensions}

We also provide three extensions, all evaluated empirically, that open the door for future work in this area.

\subsection{Empirical calculation of the risk}
\label{sec:empirical_risk}

In Section~\ref{sec:var_D} (and Appendix~\ref{sec:appendix_var_D} in more detail) we derived an upper bound on $\textrm{Var}(D)$. To help understand the risk in acquiring E\&M capabilities, here we perform an empirical calculation on $\textrm{Var}(D)$ to determine how far we are from the bound in general.

Similar to the previous experiment, for each run we randomly sample $N$, $M$, $\mu_X$, $\mu_\epsilon$, $\sigma^2_X$, $\sigma^2_1$, $\sigma^2_2$, where the restrictions $M < N$ and $\sigma^2_2 < \sigma^2_1$ are maintained, and perform 1,000 cycles of the six steps mentioned above to obtain samples of \texttt{D}. This is followed by 500 bootstrap resamplings on the samples to obtain an empirical bootstrap distribution for the variance. We then take the mean of the bootstrap variance estimates, and compare them with the theoretical upper bound by computing the ratio between the empirical variance and the bound.

\begin{figure}
\vspace*{0pt}
\begin{center}
  \includegraphics[width=0.38\textwidth, trim=2.5mm 0 1.5mm 0, clip]{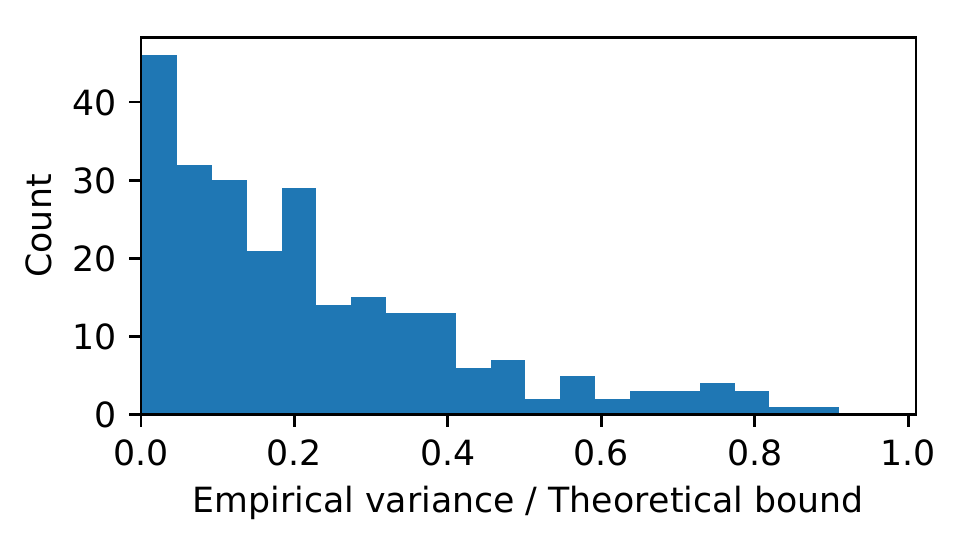}
\end{center}
\vspace*{-8pt}
\caption{The distribution of the ratios between empirical bootstrap variance estimate and the theoretical upper bound.}
\label{fig:ratio_empirical_bound_distribution}
\end{figure}

We performed 250 runs and show the distribution of ratios between the empirical variance and the bound in Figure~\ref{fig:ratio_empirical_bound_distribution}. The figure shows while most empirical values of $\textrm{Var}(D)$ are much lower than the upper bound, with around 85\% of the samples having a ratio $\leq 40\%$, in a few scenarios, the bootstrap variance estimate is up to 90\% of the upper bound, indicating a lower magnitude of the covariance term in~\eqref{eq:appendix_var_D}.

In future work it would be interesting to establish a tighter bound for, or an accurate estimate of the variance of the value gained from prioritization under different parameter combinations.

\subsection{Valuation under independent $t$-distributed assumptions}

The model described in Section~\ref{sec:normal_normal_model} assumes that both the true value of the propositions and the E\&M noise are normally distributed. While possessing decent mathematical properties, it is insufficient to explain the heavy tail in the distribution of uplifts shown in~\cite{browne17whatworksin} or~\cite{johnson17theonlinedisplay}.

In this section, we model the true value of the propositions, as well as the estimation noise, as Generalized Student's $t$-distributions:\footnote{A generalized Student's $t$-distribution is specified as $X = \mu + \sigma T_{\nu}$, where~$T_{\nu}$ is a standard Student's $t$-distribution with $\nu$ degrees of freedom, and~$\mu$ and~$\sigma$ are the location and scaling parameter respectively.
The idea is similar to `generalizing' a standard normal distribution by multiplying it with a scaling parameter and adding a location parameter.}
\begin{align}
    X_n \overset{\textrm{i.i.d}}{\sim} t_\nu(\mu_X, \sigma^2_X), \;
    \epsilon_n \overset{\textrm{i.i.d}}{\sim} t_\nu(\mu_\epsilon, \sigma^2_\epsilon) ,
    \label{eq:appendix_t_t_model}
\end{align}
where $\epsilon_n \perp X_m\, \forall n,m$, $\nu$ denotes the degrees of freedom of the underlying standard Student's $t$-distribution, $\mu$s denotes the location parameter, and $\sigma^2$s denotes the scale parameter.

It is difficult to derive theoretical quantities under such model assumptions because Student's $t$-distributions do not have conjugate priors (see e.g.~\cite{robert07thebayesian}). We instead simulate the empirical distribution of the value gained under different parameter combinations to understand if this model is a better alternative to that under normal assumptions. The sampling procedure is similar to that described Section~\ref{sec:experiments}, with Steps~1 and~2 modified such that the samples are generated from standard $t$-distributions, then scaled and located as specified by Expression~\eqref{eq:appendix_t_t_model}.

We compare the value gain estimates obtained under $t$-distributed assumptions and normal assumptions as follows. For each run, we randomly sample values for $N$, $M$, $\mu_X$, $\mu_\epsilon$, $\sigma^2_X$, $\sigma^2_1$, $\sigma^2_2$, and perform 1,000 cycles of the six-step sampling procedure Section~\ref{sec:experiments} above to obtain samples of \texttt{D} using both the $t_3$ and normal distributions.\footnote{$t_3$ ($t$-distribution with three degrees of freedom (d.f.)) is used as it is the distribution with the longest tail under the $t$ family with a natural number d.f. while retaining a finite variance.} We then compare the expected values, as well as the 5\% and 95\% percentiles of the value gained under the two distributions.

We observed from 840 runs that the distribution of value gained under the $t$-distributed assumptions has a higher mean (30\% higher on average) and variance (40\% higher in the 95\% percentile on average) than under normal assumptions, reflecting the larger spread in true value and estimation noise. Moreover, if we scale the initial $t$-distributions by $\sqrt{(\nu-2)/\nu}$ such that it has the same variance as the normal distributions, the observation still holds, albeit with a lower magnitude (7\% higher mean and 7\% higher 95\% percentile on average). This shows that the model under $t$-distributed assumptions is able to capture the ``higher risk, higher reward'' concept.

\subsection{Partial estimation / measurement noise reduction}

There are many situations when not all propositions are immediately measurable upon the acquisition of E\&M capabilities. This may be due to the extra work required to integrate additional capabilities in certain legacy systems, or the limited ability to run experiments on online but not offline activities.
In the case where there is a single backlog, we ask the question, will an organization still benefit from a partial noise reduction when some propositions' values are obtained under reduced uncertainty while others are subject to the original noise level?

We address this by attempting to establish the relationship between the expected improvement in mean true value of the selected propositions and the proportion of propositions that benefited from a reduced estimation  noise (denoted $p \in [0, 1]$).
The sampling procedure is similar to that described in Section~\ref{sec:experiments}, with Step~5 modified: when we repeat Step~2, instead of generating all samples from $\mathcal{N}(\mu_\epsilon, \sigma^2_2)$ we generate~$p$ of the samples from $\mathcal{N}(\mu_\epsilon, \sigma^2_2)$ (the lowered estimation / measurement noise) and $1-p$ of the samples from $\mathcal{N}(\mu_\epsilon, \sigma^2_1)$ (the original estimation / measurement noise).

\begin{figure*}
\begin{center}
\begin{subfigure}[t]{.11\textwidth}
  \includegraphics[width=\textwidth, trim=3.5mm 0 2.8mm 0, clip]{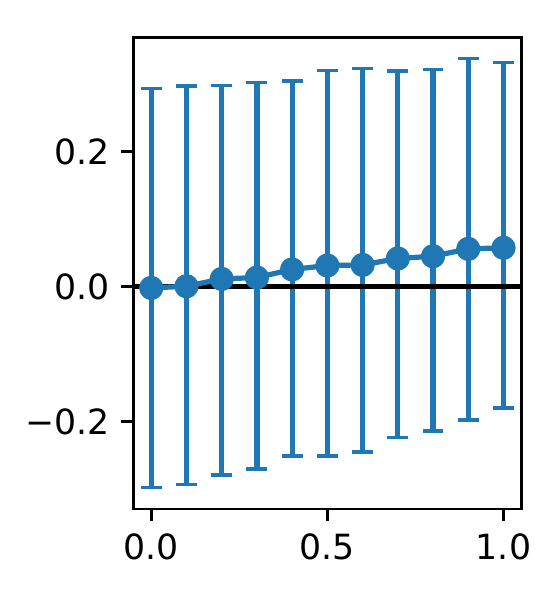}
  \caption{$N = 50$, $M = 5$}
  \label{fig:partial_noise_50_5_1_0-5_0-4}
\end{subfigure}
\hspace*{.001\textwidth}
\begin{subfigure}[t]{.115\textwidth}
  \includegraphics[width=\textwidth, trim=3.5mm 0 2.8mm 0, clip]{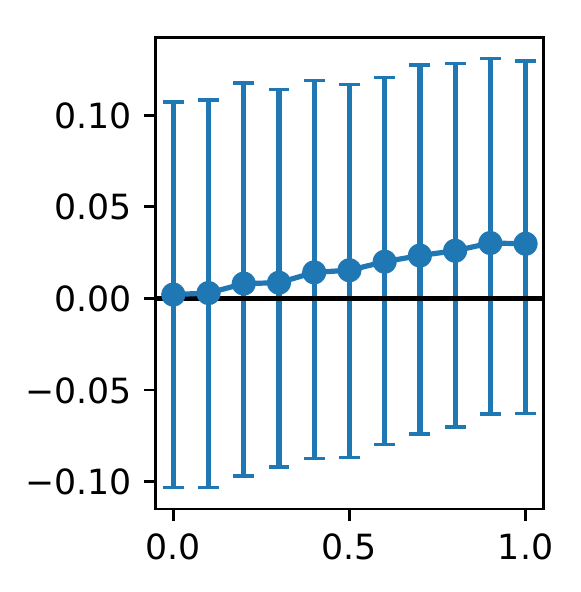}
  \caption{$N = 50$, $M = 20$}
  \label{fig:partial_noise_50_20_1_0-5_0-4}
\end{subfigure}
\hspace*{.001\textwidth}
\begin{subfigure}[t]{.11\textwidth}
  \includegraphics[width=\textwidth, trim=2.8mm 0 2.8mm 0, clip]{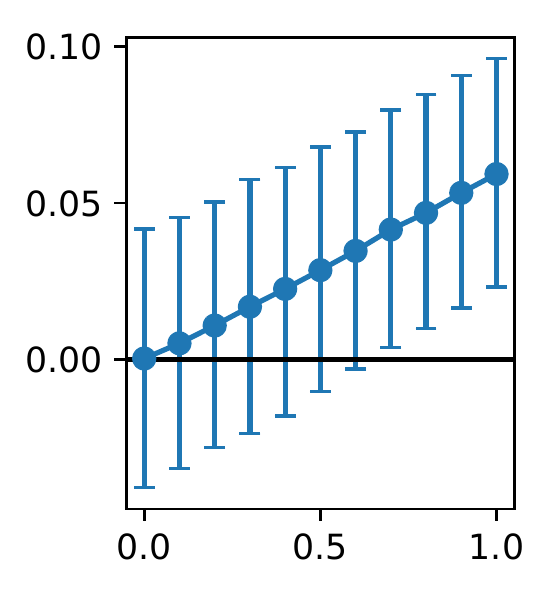}
  \caption{$N = 2500$, $M = 250$}
  \label{fig:partial_noise_2500_250_1_0-5_0-4}
\end{subfigure}
\hspace*{.001\textwidth}
\begin{subfigure}[t]{.11\textwidth}
  \includegraphics[width=\textwidth, trim=2.8mm 0 2.8mm 0, clip]{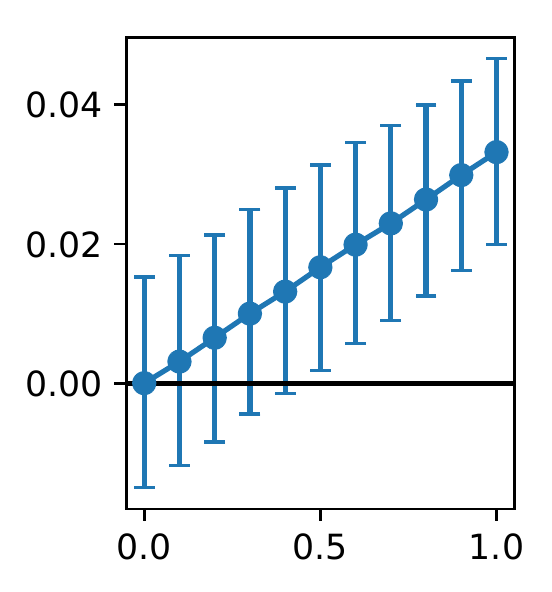}
  \caption{$N = 2500$, $M = 1000$}
  \label{fig:partial_noise_2500_1000_1_0-5_0-4}
\end{subfigure}
\hspace*{.001\textwidth}
\begin{subfigure}[t]{.115\textwidth}
  \includegraphics[width=\textwidth, trim=3.5mm 0 2.8mm 0, clip]{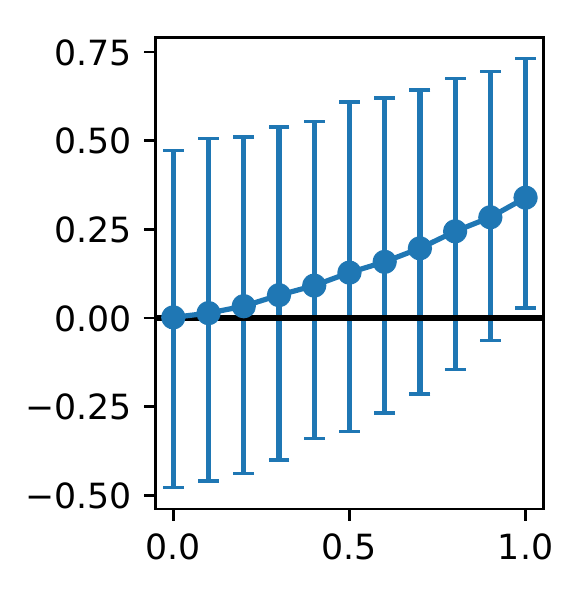}
  \caption{$N = 50$, $M = 5$}
  \label{fig:partial_noise_50_5_1_0-8_0-2}
\end{subfigure}
\hspace*{.001\textwidth}
\begin{subfigure}[t]{.11\textwidth}
  \includegraphics[width=\textwidth, trim=3.5mm 0 2.8mm 0, clip]{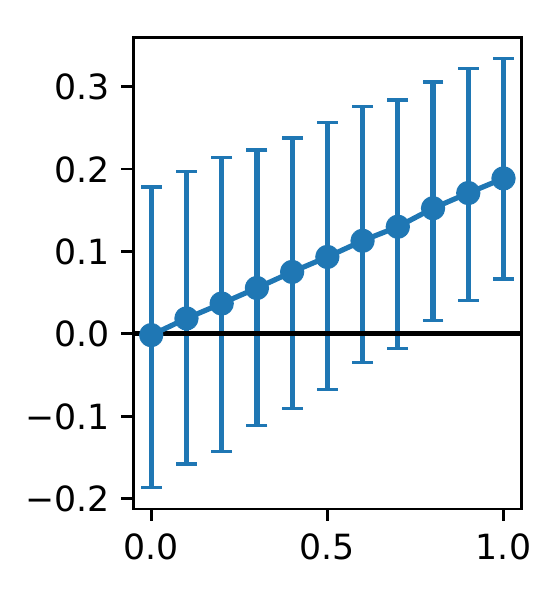}
  \caption{$N = 50$, $M = 20$}
  \label{fig:partial_noise_50_20_1_0-8_0-2}
\end{subfigure}
\hspace*{.001\textwidth}
\begin{subfigure}[t]{.105\textwidth}
  \includegraphics[width=\textwidth, trim=2.8mm 0 2.8mm 0, clip]{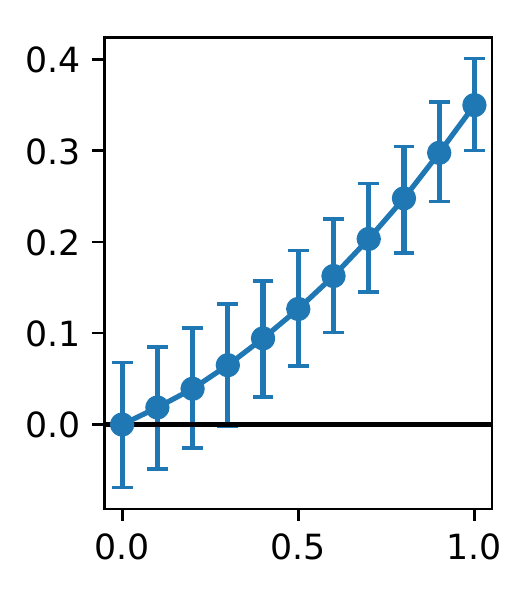}
  \caption{$N = 2500$, $M = 250$.}
  \label{fig:partial_noise_2500_250_1_0-8_0-2}
\end{subfigure}
\hspace*{.001\textwidth}
\begin{subfigure}[t]{.11\textwidth}
  \includegraphics[width=\textwidth, trim=2.8mm 0 2.8mm 0, clip]{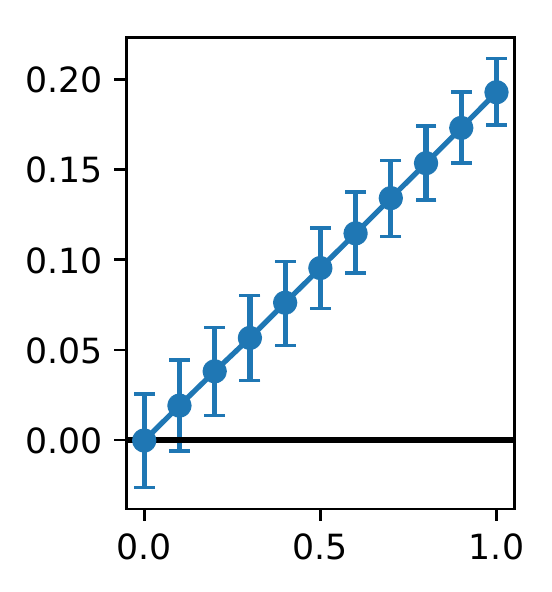}
  \caption{$N = 2500$, $M = 1000$}
  \label{fig:partial_noise_2500_1000_1_0-8_0-2}
\end{subfigure}
\end{center}
\caption{The near-linear relationship between $p$ (proportion of propositions which value is obtained under a lower estimation / measurement noise, $x$-axes) and the improvement in mean true value of the selected propositions ($y$-axes) under the normal value / normal noise model. In each plot the dot represents the sample mean, and the error bar represents the 5\% and 95\% percentile of the sample value gained (see Equation~\eqref{eq:appendix_expected_mean_true_value_diff}). All figures assume $\sigma^2_X = 1$, while the left four figures assume $\sigma^2_1 = 0.5^2$ and $\sigma^2_2 = 0.4^2$ (corresponding to a small reduction in estimation / measurement noise), and the right four figures assume $\sigma^2_1 = 0.8^2$ and $\sigma^2_2 = 0.2^2$ (corresponding to a large reduction in estimation / measurement noise).}
\label{fig:partial_noise_experiment}
\end{figure*}

We run the procedure above under various scenarios, including under a large/small ($N$), a large/small ratio between an organizations' capacity and backlog (${M}/{N}$), and a large/small magnitude of noise reduction upon acquisition of E\&M capabilities ($\sigma^2_1 - \sigma^2_2$).
Figure~\ref{fig:partial_noise_experiment} shows the result. We can see that under most scenarios, the expected value gained increases with $p$ at least linearly, while there are a few scenarios where the expected improvement in mean true value of the selected propositions curve upwards for increasing~$p$. This shows that while there are incentives for organizations to acquire E\&M capabilities that cover the majority of their work, in many scenarios, a partial acquisition yields proportional benefits. Potential experimenters need not see the acquisition as a zero-one decision, or worry about any steep initial investment required to unlock returns.

\end{appendices}

\end{document}